\documentclass[a4paper,11pt]{article}
\usepackage[utf8]{inputenc}
\usepackage[english]{babel}
\usepackage{authblk}
\usepackage{braket}
\usepackage{xspace}
\usepackage{bbm}
\usepackage{mathdots}
\usepackage{stackrel}
\usepackage[dvipsnames]{xcolor}
\usepackage{appendix}
\usepackage{hyperref}
\hypersetup{
 colorlinks=true,
 linkcolor=blue,
 anchorcolor = blue,
 citecolor = blue,
 filecolor = blue,
 urlcolor = blue
}
\usepackage{mathtools}
\usepackage{bbm}
\usepackage{comment}
\usepackage{subfigure}
\usepackage[T1]{fontenc}               % codifica dei font
\usepackage[left=3cm,
            right=2.5cm,
            top=2.5cm,
            bottom=3cm
            ]{geometry}                % spaziatura bordi
\usepackage{graphicx}
\usepackage{mathrsfs}
\usepackage{appendix}
\usepackage{fancyhdr}
\usepackage{amsmath,                   % simboli matematici
            amssymb,                   % altri simboli matematici
            amsthm}                    % stili teorema

\usepackage{setspace}
\usepackage{cite}
\usepackage{cancel}
\usepackage{bm}
\usepackage[margin=30pt, bf, font=small, center, justification=justified]{caption}[2004/07/16]

\usepackage{setspace} 
\usepackage{tikz}

\newcommand{\trho}{\tilde{\rho}}

\newcommand{\exB}{\mathbb{E}_{\text{Born}}}

\title{Gaussian Scrooge Ensemble from Deep Thermalization in Free-Fermions}

\author[1]{Angelo Russotto}
\author[2]{Katja Klobas}
\author[1]{Pasquale Calabrese}
\author[2]{Bruno Bertini}

\affil[1]{SISSA and INFN, via Bonomea 265, 34136 Trieste, Italy}
\affil[2]{School of Physics and Astronomy, University of Birmingham, Edgbaston, Birmingham, B15 2TT, UK}

\date{}
\begin{document} 

\maketitle

%{\small
%\vspace{-5mm}  \ \\
%{$^{1}$}  SISSA and INFN, via Bonomea 265, 34136 Trieste, Italy\\[-0.1cm]
%\medskip
%%{$^{2}$} School of Physics and Astronomy, University of Birmingham, Edgbaston, Birmingham, B15 2TT, UK
%\medskip
%}

\begin{abstract}
Deep thermalization indicates the emergence of universal ensembles of pure states in a subsystem after measuring its environment. We investigate this phenomenon in a free-fermion chain using local measurements that preserve Gaussianity and break particle-number conservation. Although the measurement outcomes provide partial information about local conserved charges within a finite subsystem, we argue that this information becomes negligible in the thermodynamic limit. This effective non-revelation underlies the emergence of a fermionic Gaussian Scrooge ensemble, determined by the subsystem's generalized Gibbs ensemble. We construct this universal ensemble directly at the subsystem level and provide numerical evidence that it describes the late-time projected ensemble at finite temperature. Our results extend the description of thermalization in non-interacting integrable systems beyond the reduced density matrix to universal statistics of conditional pure states.
\end{abstract}

\tableofcontents

\section{Introduction}

Understanding how equilibrium emerges from unitary quantum dynamics is a central problem in many-body physics. Conventional thermalization concerns the relaxation of local observables and is described by the reduced density matrix of a subsystem much smaller than its surrounding bath. In generic systems, the stationary states are locally described by Gibbs ensembles~\cite{d-91,s-94,s-99,rdo-08,polkov-11,dalessio-16,d-18}. In integrable systems, an extensive set of additional conserved quantities constrains relaxation, leading instead to a generalized Gibbs ensemble (GGE)~\cite{rigol-07,cef-11,ef-13,ilievski-15,ilievski-16,vr-16,pvc-16,piroli-17,sbkc-26}. These descriptions determine local expectation values, but do not address the additional information obtained by measuring the surrounding system and recording the outcome, which is now a standard feature of real-world quantum simulators. To take into account this additional information it is useful to consider the \emph{projected ensemble}: an ensemble of pure states conditioned on measurements of the partial state, weighted by their Born probability. The approach of this ensemble to universal distributions is known as the \emph{deep thermalisation}.

A recent surge of interest on this phenomenon has been triggered by experimental observations in a Rydberg-atom-array quantum simulator~\cite{choi-23}, together with theoretical studies~\cite{cotler-23,hc-22}. Subsequent experiments have explored universal quantum-state statistics in several platforms~\cite{shaw-25,yan-26,liu-26}. Numerous theoretical studies have characterized deep thermalization in quantum circuits~\cite{hc-22,cl-22,ih-22,ih-23,msr-26}, Hamiltonian systems~\cite{cotler-23,mark-24,sh-25,mok-26}, random tensor networks~\cite{lami-26}, Gaussian bosonic~\cite{lhh-24} and fermionic systems~\cite{lpdd-23,bbm-25}, as well as extensions to mixed-state ensembles~\cite{yhk-25,mm-25,sr-26}, and instances in which it fails~\cite{bdp-23,lih-26,feng-26}. Beyond its fundamental significance, deep thermalization provides a source of quantum randomness for applications in quantum information~\cite{mok-25}, including benchmarking~\cite{choi-23,mark-23}, state learning~\cite{tran-23}, and quantum cryptography~\cite{chakraborty-25,cui-25} (see Ref.~\cite{mark-26} for a recent review).

A particularly interesting aspect of deep thermalization is the interplay between conservation laws and measurements~\cite{vb-24,chang-25}. Conserved quantities constrain the stationary reduced density matrix, while measurements of the environment can reveal additional information about the charges in the subsystem. In chaotic systems, measurements that reveal no additional charge information are expected to produce a Scrooge ensemble~\cite{mark-24,chang-25,mok-26,ms-25}, a Haar distribution distorted by the presence of the conservation laws~\cite{jrw-94,goldstein-06,goldstein-16}. Measurements that reveal charge information can instead lead to mixtures of Scrooge ensembles~\cite{chang-25}. Understanding which information survives in the thermodynamic limit is therefore essential for identifying the universal projected ensemble.

Integrable systems offer a particularly useful setting in which to address this question because of their many conservation laws, while free-fermion models permit the study of large systems through correlation matrix methods whose computational cost is only polynomial in the total system size. Previous work on free fermions introduced the \emph{deep generalized Gibbs ensemble} as an ansatz for the projected ensemble. The latter is constructed by measuring a representative global state with the appropriate conserved-charge densities~\cite{lpdd-23}. This construction, however, requires knowledge of the full state. Related work rigorously established convergence to a uniform ensemble of fermionic Gaussian states in random matchgate circuits in the absence of conserved charges~\cite{bbm-25}, while Gaussian Scrooge distributions were identified in bosonic continuous-variable systems~\cite{lhh-24}. These results motivate a description of the universal finite-temperature projected ensemble of Gaussian fermions directly on the subsystem.

In this work, we address this problem by investigating a Gaussian integrable Hamiltonian with a large number of conserved charges that are only partially accessible or revealed by the bath measurements. We examine how the information revealed about local conserved charges affects a fixed subsystem as the bath grows. Our analysis shows that the partial information about conserved charges becomes irrelevant in the thermodynamic limit, which motivates a universal description in terms of a fermionic Gaussian Scrooge ensemble constructed directly from the reduced $\rm{GGE}$.

The manuscript is organized as follows. In Sec.~\ref{sec:DT} we briefly review the concept of deep thermalization and discuss the state of the art. In Sec.~\ref{sec:model} we introduce the model and the measurement protocol analyzed in the rest of the manuscript. In Sec.~\ref{sec:inftemp} we first consider the simpler case of quenches leading to a locally infinite-temperature state, which allows us to highlight the asymptotic non-revelation of the conserved charges. In Sec.~\ref{sec:finittemp}, we generalize our predictions to finite-temperature quenches by introducing the fermionic Gaussian Scrooge ensemble. Finally, we present our conclusions in Sec.~\ref{sec:concl}. Further technical details and derivations are provided in the appendices.

\begin{figure*}[t]
\centering
\includegraphics[width=0.79\textwidth]{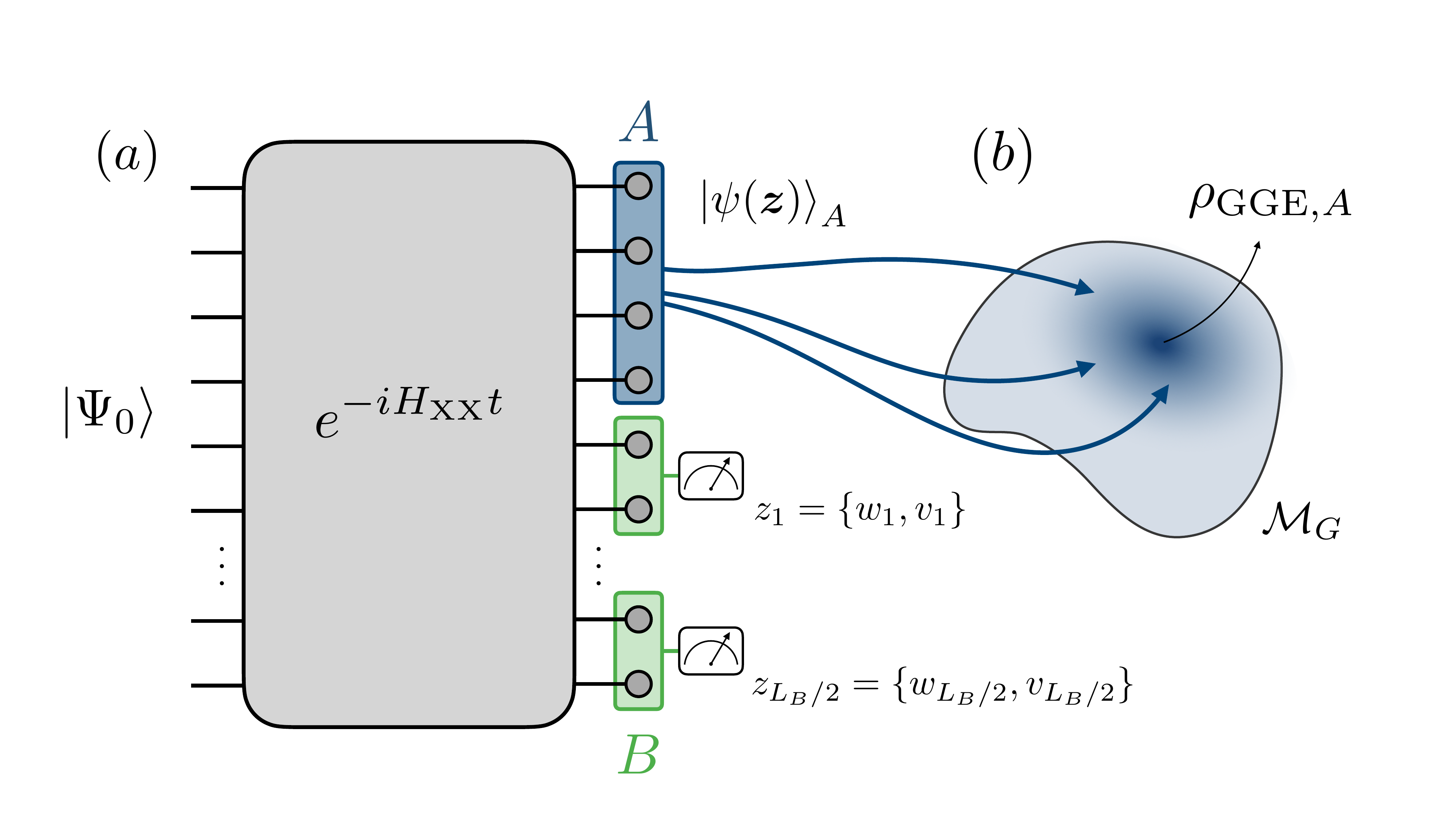}
    \caption{\label{fig:sketch} Deep thermalization in a fermionic Gaussian  quantum many-body system. (a) An initial Gaussian state $\ket{\Psi_0}$ is time-evolved with the $\rm{XX}$ Hamiltonian \eqref{eq:Hxx}. At time $t$ the bath $B$ is projectively measured with outcome $\bm{z} = \{z_1,\cdots,z_{L_B/2}\}$. The measurement basis is chosen to break the infinitely many conservation laws of the $\rm{XX}$ Hamiltonian. (b) The ensemble of post-measurement pure states $\ket{\psi(\bm{z})}_A$ in $A$ approaches, in the thermodynamic limit $L_B\to\infty$ and then $t\to \infty$, a Gaussian Scrooge ensemble, a uniform distribution over the manifold $\mathcal{M}_G$ of Gaussian state, distorted such that the first moment is the reduced Generalized Gibbs Ensemble $\rho_{\mathrm{GGE},A}$.} 
\end{figure*}

\section{Deep Thermalization}
\label{sec:DT}
In this section, we introduce the notion of projected ensemble and the idea underlying deep thermalization. Consider a pure state $\ket{\Psi}$ defined on a bipartite system $A\cup B$, where $A$ and $B$ respectively contain $L_A$ and $L_B$  qubits. In the following we identify $A$ as our subsystem of interest and $B$ with the bath and according consider $L_B \gg L_A$. Suppose now to projectively measure the bath $B$ in such a way as to fully determine the state in $B$, i.e.\ with a rank-1 projection. Without loss of generality, we can encode the outcome of the measurement in a binary string $\bm{z} \in \{0,1\}^{L_B}$. To each outcome $\bm{z}$, occurring with Born probability $p(\bm{z}) = \bra{\Psi}(\mathbb{I}_A \otimes \ket{\bm{z}}\!\bra{\bm{z}})\ket{\Psi}$, we can associate the projected pure state $\ket{\psi(\bm{z})}_A =(\mathbb{I}_A \otimes \bra{\bm{z}}) \ket{\Psi} /\sqrt{p(\bm{z})}$ in $A$. The projected ensemble $\mathcal{E}_{\rm{PE}}$ is a distribution of pure states in $A$ defined as the collection of projected states $\ket{\psi(\bm{z})}_A$ each weighted by its Born probability $p(\bm{z})$
\begin{equation}
\label{eq:def_PE}
\mathcal{E}_{\text{PE}} = \{p(\bm{z}), \ket{\psi(\bm{z})}_A\}.
\end{equation}
The first moment of $\mathcal{E}_{\rm{PE}}$ coincides with the partial trace over the bath, i.e.
\begin{equation}
 \rho_{\mathcal{E}_{\rm{PE}}}^{(1)}= \mathbb{E}_{\ket{\psi} \sim \mathcal{E}_{\rm{PE}}}[\ket{\psi}\!\bra{\psi}] = \text{Tr}_B[\ket{\Psi} \bra{\Psi}]=\rho_A,
\end{equation}
while higher moments of the projected ensemble
\begin{equation}  \rho_{\mathcal{E}_{\rm{PE}}}^{(k)} = \sum_{\bm{z}}p(\bm{z}) \left(\ket{\psi(\bm{z})}\!\bra{\psi(\bm{z})}\right)^{\otimes k},
\end{equation}
encode the statistics of conditional expectation values $\langle \mathcal{O}_{A}\rangle_{\bm{z}}$ of local operators in $A$ through
\begin{equation}
    \mathbb{E}_{\ket{\psi} \sim \mathcal{E}_{\rm{PE}}} \left[ \langle \mathcal{O}_{A}\rangle^k_{\bm{z}}\right] = \text{Tr}\left[\rho_{\mathcal{E}_{\rm{PE}}}^{(k)}\mathcal{O}_A^{\otimes k}\right].
\end{equation}

In isolated many-body systems, such as the one considered in this work, conventional thermalization is formulated in terms of the emergence of a local equilibrium state $\rho_{\mathrm{eq},A}$ so that 
\begin{equation}
  \lim_{t\to\infty}\lim_{L_{B}\to\infty} \bra{\Psi(t)}\mathcal{O}_A \ket{\Psi(t)} 
  =\text{Tr}_A[\rho_{\mathrm{eq},A} \mathcal{O}_A],
\end{equation}
which can be equivalently recast as a constraint on the first moment of the projected ensemble 
\begin{equation}
  \lim_{t\to\infty}\lim_{L_B\to\infty}\rho^{(1)}_{\mathcal{E}_{\rm{PE}}} 
  = \rho_{\mathrm{eq},A}.
\end{equation}
Deep thermalization extends this notion by asking whether the full hierarchy of higher moments, and therefore the full distribution of conditional pure states, also approach a universal form.

In the absence of conservation laws, the expected universal ensemble is the Haar ensemble of pure states~\cite{cotler-23,hc-22,ih-23,cl-22,ih-22}. Instead, in the presence of conservation laws, the local equilibrium state is generally not maximally mixed but instead retains information about the conserved quantities. Moreover, unlike in the conventional thermalization, the limiting projected ensemble is constrained not only by the conservation laws, but also by the measurement basis in the bath $B$ \cite{mark-24,vb-24,chang-25}. The effect of the measurement basis can be classified according to the amount of information that the measurement outcomes reveal about the conserved charge in $A$. Consider, for simplicity, a local $U(1)$ charge $Q = Q_A + Q_B$. Depending on the measurement basis, a certain outcome $\bm{z}$ can update the knowledge of the charge $Q_A$.  This motivates a general classification of measurement bases \cite{chang-25}, ranging from \textit{maximally charge revealing} basis, in which $\bm{z}$ uniquely determines the value of $Q_B$, to \textit{charge non-revealing} basis, in which conditioning on the measurement outcome leaves the subsystem charge distribution unchanged, see Sec.~\ref{subsec:nonrevel} for a more precise discussion.

For charge non-revealing measurement bases, the expected universal projected ensemble is the Scrooge ensemble~\cite{mark-24,chang-25} associated with the local equilibrium state. The Scrooge ensemble associated with a density matrix $\rho$ consists of the normalized states~\cite{jrw-94,goldstein-06,goldstein-16}
\begin{equation}
|\psi(\phi)\rangle
=\frac{\sqrt{\rho}\,|\phi\rangle}
{\sqrt{\langle\phi|\rho|\phi\rangle}},
\end{equation}
where $|\phi\rangle$ is sampled from the Haar measure with probability density $D\langle\phi|\rho|\phi\rangle$, and $D$ is the Hilbert-space dimension. This ensemble has average density matrix $\rho$ and reduces to the Haar ensemble when $\rho=\mathbb I/D$.
In particular, this prediction has been supported numerically for nonintegrable, energy-conserving many-body systems where the bath is measured in an energy-non-revealing basis~\cite{cotler-23,mark-24}. The underlying principle behind the emergence of Scrooge ensembles is the minimization of the \textit{accessible information}~\cite{holevo-98} or, equivalently, maximization of ensemble entropy over the accessible Hilbert space~\cite{mark-24}. More broadly, Scrooge ensembles also provides a framework for canonical and dynamical typicality~\cite{ttv-25,psz-26}.

More generally, bath measurements can partially reveal the conserved charge. In the $U(1)$ setting described above, outcomes can be grouped into equivalence classes according to the conditional charge distribution $p(Q_A|\bm{z})$, defined using an appropriate ensemble of representative global states~\cite{chang-25}. Each class can then be associated with a Scrooge ensemble, giving rise to a mixture known as the generalized Scrooge ensemble~\cite{mark-24,chang-25}. 
The latter case is of particular relevance for Hamiltonian dynamics, in which, typically, energy does not admit a local fully revealing basis. For a generic measurement basis, explicitly specifying the mixture can require an amount of data that grows exponentially with bath size~\cite{mark-24}, raising questions on the degree of universality of the Scrooge ensemble. Understanding the interplay between the measurement basis and the conservation of generic finite-range local charges is thus crucial to determine to what extent the projected ensemble deeply thermalises to a truly universal ensemble.

\section{Model and measurement protocol}
\label{sec:model}

We consider a one-dimensional chain of spinless fermions on an even number $L$ of sites. The system is prepared in low-entangled, number-conserving, pure, Gaussian state $\ket{\Psi_0}$ and evolves in time with a unitary dynamics described by the quadratic Hamiltonian 
\begin{equation}
\label{eq:Hxx}
H_{\rm{XX}}= - \sum_{j=1}^L( c^{\dagger}_{j+1}c_j + c^{\dagger}_j c_{j+1}),
\end{equation}
where we assume periodic boundary conditions $c_{L+1} = c_1$. We consider a bipartition of the system into a subsystem $A$ corresponding to sites $j = 1,\cdots, L_A$ with $L_A$ even and the bath $B$ with the remaining $L_B = L-L_A$ sites. The model in Eq.~\eqref{eq:Hxx} is integrable with an extensive set of local conserved charges~\cite{fagotti-14,ef-16}, therefore, in the limit of $L_B \to \infty$ and $t \to \infty$ the stationary state in $A$ is described by the reduced GGE\footnote{Note that, since the initial state has fixed particle number, we do not need to consider the extra non-abelian charges in Eq.~\eqref{eq:xxcharges} and can describe the GGE solely in terms of occupation numbers.} 
\begin{equation}
    \rho_{\mathrm{GGE},A} = \frac{1}{Z} \text{Tr}_B\left[e^{- \sum_k \lambda(k)\, n_k} \right],
\end{equation}
with 
\begin{equation}
n_k = c^{\dagger}_k c_k, 
\quad\text{and}\quad
c_k = \frac{1}{\sqrt{L}} \sum_j e^{-ikj} c_j. 
\end{equation}
The Lagrange multipliers $\lambda(k) = \log((1-n(k))/{n(k)})$ are related to the momentum density occupation function $n(k) = \text{Tr}[\rho_0 n_k]$ of the initial state $\rho_0 = \ket{\Psi_0}\!\bra{\Psi_0}$. At any time $t$ of the quantum quench the state $\ket{\Psi(t)} = e^{- i H_{\rm{XX}}t} \ket{\Psi_0}$ is Gaussian, therefore all physical observables are encoded in the two-point correlation matrix $C_{ij} = \bra{\Psi} c^{\dagger}_i c_j \ket{\Psi}$. 

In the following, we will conveniently describe the system in terms of the $2 L$ Majorana fermions
\begin{equation}
\gamma_{j,1} = c_j + c^{\dagger}_j, \qquad  \gamma_{j,2} = i(c_j - c^{\dagger}_j),
\end{equation}
which satisfy the algebra $\{\gamma_{j,\alpha},\gamma_{k,\beta}  \} = 2 \delta_{jk} \delta_{\alpha \beta}$. Introducing the composite index $a = (j,\alpha)$, the Majorana correlation matrix $\Gamma$ of a state $\ket{\Psi}$ is defined as the real anti-symmetric matrix 
\begin{equation}
\Gamma_{ab} =i( \bra{\Psi} \gamma_{a} \gamma_b \ket{\Psi}- \delta_{ab}).
\end{equation}

\noindent \textbf{Measurement protocol.} Our goal is to projectively measure the bath $B$ such that, after the measurements, the projected state in $A$ is pure, Gaussian and, importantly, such that it breaks particle-number conservation, with the objective of having a particle number non-revealing measurement basis. 

We can achieve the required properties in the following way. We divide $B$ into $ (L-L_A)/2$ two-site cells and measure in each cell the following two Hermitian operators $M_1$ and $M_2$ (see Fig.~\ref{fig:sketch} for a sketch) acting on sites $2j-1,2j$
\begin{equation}\label{eq:defM12}
  \begin{aligned}
    M_{j,1} &\equiv  i \gamma_{2j-1,1} \gamma_{2j,1} 
    = i \left(c^{\dagger}_{2j-1} c_{2j} + c^{\dagger}_{2j-1} c^{\dagger}_{2j} - \mathrm{h.c.}\right),\\
    M_{j,2} &\equiv  i \gamma_{2j-1,2} \gamma_{2j,2} 
    = i \left(c^{\dagger}_{2j-1} c_{2j} - c^{\dagger}_{2j-1} c^{\dagger}_{2j} - \mathrm{h.c.}\right).
  \end{aligned}
\end{equation}
These two operators commute $\left[M_{j,1},M_{k,2}\right] = 0$ and are not proportional to each other, which implies that measuring them sequentially provides a rank-1 projection on the two-site cell. Moreover, since $M_{j,1}^2 = M_{j,2}^2 = 1$, we can write the projectors associated with different measurement outcomes $z = \pm 1$ as 
\begin{equation}
\label{eq:Pialpha}
\Pi^{(\alpha)}_{z} = \frac{\mathbb{I} + z M_{\alpha}}{2} \qquad \alpha = 1,2, 
\end{equation}
which preserves Gaussianity, as can be shown by writing each projector as
\begin{equation}
\frac{\mathbb{I} + z M_{\alpha}}{2} 
  = \lim_{\mu \to \infty} \frac{2\ e^{\mu z M_{\alpha}}}{\text{Tr}[e^{\mu z M_{\alpha}}]},
  %e^{\mu z M_{\alpha}},
\end{equation}
with $M_{\alpha}$ a quadratic operator by definition. Observe that, in spin language, the measurement operators in Eq.~\eqref{eq:defM12} correspond to $M_{j,1} = Y_{2j-1} X_{2j}$ and $M_{j,2} = - X_{2j-1} Y_{2j}$, therefore, apart from an $i$ phase to the $2j$-qubit, the projected states in the two-site cell are Bell basis states~\cite{nc-10}.

Given two outcomes $w = \pm 1$ and $v = \pm 1$, respectively for $M_1$ and $M_2$, $\Pi_{v}^{(2)}\Pi_{w}^{(1)}$ is a rank-1 projector on a two-qubit Hilbert space. Since 
\begin{equation}
[M_{1,2},n_{2j-1}+ n_{2j}] \neq 0,
\end{equation}
where we set 
\begin{equation}
n_j=c^\dag_j c^{\phantom\dag}_j,
\end{equation}
the projected state breaks particle-number conservation. Performing the measurements for each disjoint two-site cell in the bath $B$, we obtain a pure Gaussian state in $A$.  We denote by $\bm{w} = \{w_i\}_{i = 1,\cdots,L_B/2}$ the collection of outcomes of the projective measurement of $M_{1}$ in the two-site cells of $B$, while we denote by $\bm{v} = \{v_i\}_{i = 1,\cdots,L_B/2}$ the outcomes for $M_2$. The total collection of outcomes $\bm{z} = \{\bm{w},\bm{v}\}$ defines the projected pure state $\ket{\psi(\bm{z})}_A$ of $A$ belonging to the projected ensemble $\mathcal{E}_{\rm{PE}}$ (cf.\ Eq.~\eqref{eq:def_PE}). Since $\mathcal{E}_{\rm{PE}}$ is an ensemble of Gaussian states, we can fully characterize it through the statistical moments of the ensemble of projected correlation matrices $\Gamma(\bm{z})$, the $k-$th moment is
\begin{equation}
\label{eq:PEmom}
\Gamma^{(k)}_{\rm{PE}} = \sum_{\bm{z}} p(\bm{z}) \,\Gamma(\bm{z})^{\otimes k}.
\end{equation}
In App.~\ref{app:app_Gammaz}, we describe how to analytically determine the correlation matrix $\Gamma(\bm{z})$ of the post-measurement Gaussian state in $A$.\\

\noindent \textbf{General family of Gaussian measurement protocols.} We have described a specific choice of the measurement operators $M_{1,2}$ in Eq.~\eqref{eq:defM12}. More generally, it is possible to construct a continuous family of measurement protocols with the same properties. To accomplish this, we can think of applying, to each two-site cell that will be monitored, the same Gaussian unitary operation $U_O = \exp(-\bm{\gamma}^T A \bm{\gamma}/4)$, with $A^T = -A$ and $e^A = O \in SO(4)$. We then measure the Majorana bilinears in Eq.~\eqref{eq:defM12}. With this procedure, by choosing $O \in SO(4)$ we can generate the full family of Gaussianity-preserving measurement protocols with measured bilinears given by $M_1^{(O)} = U_O^{\dagger} M_1 U_O$ and $M_2^{(O)} = U_O^{\dagger} M_2 U_O$ with a $SO(2) \times SO(2)$ redundancy given by independent rotations of the two Majorana operators in $M_{1,2}$ and by exchanging $M_1 \leftrightarrow M_2$. In the following, we refer to this general family of measurement protocols to discuss which physically relevant consequences are independent of the specific choice of measurement operators.

\section{Infinite-temperature projected ensemble}
\label{sec:inftemp}

We begin our analysis by considering the case of an infinite temperature quench. Hence, we require that the initial state $\ket{\Psi_0}$ satisfies the property $\bra{\Psi_0} c^{\dagger}_k c^{\phantom\dag}_k\ket{\Psi_0} = 1/2$ for any momentum $k$, which guarantees that the subsystem $A$ locally thermalizes to  the maximally mixed state $\rho_{\mathrm{GGE},A} = \mathbb{I}/2^{L_A}$. According to the logic introduced in Sec.~\ref{sec:DT}, the first step to understand the properties of the limiting projected ensemble is the analysis of the conserved charges present in the system and, crucially, of the amount of information they reveal.

\subsection{Effective non-revelation of local conserved charges}
\label{subsec:nonrevel}

We begin our analysis of the effect of the measurement basis by considering the fermionic parity, which is necessarily conserved by the dynamics. We then analyze explicitly the effect on the particle number and the two-local conserved charges present in the model in Eq.~\eqref{eq:Hxx}. \\

\noindent \textbf{Fermionic parity.} A first important observation is 
\begin{equation}
M_1 M_2 = - Z_{2j-1} Z_{2j} = - P,
\end{equation}
where $P = (-1)^{n_{2j-1}+n_{2j}}$ is the fermionic parity. This means that, if we denote by $v$ and $w$ the outcome of $M_1$ and $M_2$, respectively, the fermionic parity on the measured sites is uniquely determined by the measurement outcome. We denote the fermionic parity of the bath associated with $\bm{z}$ by $\pi_{\bm{z}}$. Note that this property is not specific to the measurement protocol chosen, as any protocol in the continuous family of measurement operators introduced in Sec.~\ref{sec:model} satisfies $M_1^{(O)} M_2^{(O)} = -P$ since $[U_O,P] = 0$ for any $O\in SO(4)$. \\

\noindent \textbf{Particle number.} Consider now the particle number 
\begin{equation}
N = \sum_{j} c^{\dagger}_j c^{\phantom\dag}_j = N_A + N_B.
\end{equation} 
Let us assume that, for a certain measurement outcome $\bm{z}$, there are $L_B/2-b_{\bm{z}}$ occurrences of $v = w$  such that the fermionic parity is odd. On such bonds, the outcome exactly fixes the number of particles on the bond to be $1$. On the remaining $b_{\bm{z}}$ bonds, because of our specific choice of measurement operators in Eq.~\eqref{eq:defM12}, the particle number is equally likely to be $0$ or $2$ with independent probabilities on different measured bonds. Hence, the exact conditional probability distribution of the particle number in $B$, given by
\begin{equation} 
\label{eq:pnbz}
  p_{\text{out}}(N_B|\bm{z}) 
  = \frac{1}{2^{b_{\bm{z}}}} \binom{b_{\bm{z}}}{\frac{N_B - (L_B/2-b_{\bm{z}})}{2}},
\end{equation}
when $N_B$ has the same parity as $L_B/2-b_{\bm{z}}$ and zero otherwise.
In particular, the conditional mean is $\langle N_B\rangle_{\bm{z}} = L_B/2$, while the conditional variance 
\begin{equation} 
\text{Var}_{\bm{z}}(N_B) \equiv \bra{\bm{z}} N_B^2 \ket{\bm{z}}- \bra{\bm{z}} N_B \ket{\bm{z}}^2 = b_{\bm{z}}.
\end{equation} 
Assuming local relaxation to infinite temperature, Chebyshev's inequality implies that (see App.~\ref{app:variance} for the details), for typical Born trajectories, the conditional number variance $\text{Var}_{\bm{z}}(N_B)$ $\simeq L_B/4$ is extensive in the bath size.

The result in Eq.~\eqref{eq:pnbz} highlights one of the main properties of our measurement protocol. Even though we have constructed the measurement to explicitly break the particle number conservation law, the bath distribution of particle number depends explicitly on the outcome $\bm{z}$, which will affect the knowledge of the subsystem distribution. According to the classification of Ref.~\cite{chang-25} described in Sec.~\ref{sec:DT}, our measurement basis is partially-revealing with respect to the particle number. We now show how the same property remains true for the higher local conserved charges.\\

\noindent \textbf{Higher-range conserved charges.} The complete set of local conserved charges for the $\rm{XX}$ model is given by \cite{fagotti-14,ef-16}
\begin{equation}\label{eq:xxcharges}
  %\begin{cases}
  \begin{aligned}
    H_r &= \sum_{j} c^{\dagger}_j c_{j+r} + c^{\dagger}_{j+r} c_j, &
    \Delta_r &= \sum_j (-1)^j (c_j c_{j+r} - c^{\dagger}_j c^{\dagger}_{j+r}),\\
    J_r &= i \sum_j (c^{\dagger}_j c_{j+r} - c^{\dagger}_{j+r} c_j),&
    \tilde{\Delta}_r &= i  \sum_j (-1)^j (c_j c_{j+r} + c^{\dagger}_j c^{\dagger}_{j+r}),
    %\end{cases}
  \end{aligned}
\end{equation}
for $r\geq 1$, together with the particle number $N = \sum_j c^{\dagger}_j c_j$. We have included the $U(1)$ breaking charges $\Delta_r, \tilde{\Delta}_r$ because of the symmetry-breaking nature of the measurements. From the definition in Eq.~\eqref{eq:xxcharges}, we can immediately observe the following relations
\begin{equation}
\label{eq:JMDM}
  J_1 = \sum_j \frac{M_{1,j}  + M_{2,j}}{2}\quad\text{and}\quad
\Delta_1 = \sum_j (-1)^j \frac{M_{1,j}  - M_{2,j}}{2}.
\end{equation}
Consider for example the $J_1$ charge. We split its contributions into 
\begin{equation}
    J_1 = J_A + J_{AB} + J_{B,e} + J_{B,o},
\end{equation}
with $J_A$ containing terms only supported in the subsystem $A$, $J_{AB}$ contains terms supported both on $A$ and $B$. The remaining terms have support only in the bath $B$, with $J_{B,e}$ containing only the terms which are measured (even bonds) and $J_{B,o}$ the terms in between the measured bonds. Analysing as before the mean and variance of the distribution $p_{\text{out}}(J_B|\bm{z})$ of the bath charge $J_B = J_{B,e} + J_{B,o}$ after a certain outcome $\bm{z}$ we obtain
\begin{equation}
    \bra{\bm{z}} J_{B,e} \ket{\bm{z}} = \sum_{j_e} \frac{v_{j_e} + w_{j_e}}{2},
\end{equation}
on the measured bonds, while
\begin{equation}
    \bra{\bm{z}} J_{B,o} \ket{\bm{z}} = 0,
\end{equation}
on the unmeasured ones. The conditional variance $\text{Var}_{\bm{z}}(J_B)$ is exactly given by 
\begin{equation}
    \text{Var}_{\bm{z}}(J_B) = \frac{L_B}{4} -\frac{1}{2},
\end{equation}
independently of the outcome $\bm{z}$. Similar result holds for the other partially revealed two-local charge $\Delta_1$. With the same approach, one can easily verify that the posterior distribution of charges with larger support is also $\bm{z}$ dependent. An important caveat in the previous analysis is the fact that the conserved charges in the $\rm{XX}$ model are not mutually commuting, and we cannot define a joint probability distribution for the charges. We therefore characterize the information revealed by the measurements through the conditional distribution of each linear combination of charges.

Any conserved charge is described by an antisymmetric matrix $X$ in the Majorana operator space $Q_X = \frac{i}{4} \bm{\gamma}^T X \bm{\gamma}$ such that $[h_{\rm{XX}},X] = 0$, where $h_{\rm{XX}}$ is the one-particle representation of the $\rm{XX}$ Hamiltonian. For a fixed subsystem $A$ with $L_A$ sites we can only consider charges with local terms supported on at most $L_A$ sites. Any linear superposition $X_{\bm{u}} = \sum_j u_j X_j$ defines a conserved charge $Q_{\bm{u}}$. The correlation matrix in the post-measurement state $\Gamma_{\bm{z}}$ splits as $\Gamma_{A,\bm{z}} \oplus \Gamma_{B,\bm{z}}$ and the bath correlation matrix is
\begin{equation}
\label{eq:GBz}
    \Gamma_{B,\bm{z}} = \sum_{\alpha=1}^{L_B} z_\alpha E_{\alpha}.
\end{equation}
In Eq.~\eqref{eq:GBz}, the sum is over the components of $z$, $z_\alpha = \pm 1$, and $E_{\alpha}$ is a $2 L_B\times 2L_B$ matrix where the only non-zero entries are $(E_{\alpha})_{2\alpha-1,2\alpha}=-(E_{\alpha})_{2\alpha,2\alpha-1}=1$. 

For a fixed outcome $\bm{z}$, the average and variance of the bath charge can be obtained using Wick's theorem 
\begin{equation} \label{eq:muzVarz}
  \begin{aligned}
    \langle Q^B_{\bm{u}}\rangle_{\bm{z}} &= 
    -\frac{1}{4} \text{Tr}[X_{\bm{u}}\,\Gamma_{B,\bm{z}}],\\
    \text{Var}_{\bm{z}}(Q^B_{\bm{u}}) &= 
    \frac{1}{16} \text{Tr}[[X_{\bm{u}},\Gamma_{B,\bm{z}}]^T [X_{\bm{u}},\Gamma_{B,\bm{z}}]] = \frac{1}{16} ||[X_{\bm{u}},\Gamma_{B,\bm{z}}] ||^2_F,
  \end{aligned}
\end{equation}
where $X_{\bm{u}}$ is considered as the restriction on $B$ of the global charge matrix.
The first consequence of Eq.~\eqref{eq:muzVarz} is the absence of perfectly revealed conserved charges, i.e.\ such that $\text{Var}_{\bm{z}}(Q^B_{\bm{u}}) = 0$ for some fixed $\bm{u}$ and for any trajectory $\bm{z}$. The necessary condition would be $[X_{\bm{u}},\Gamma_{B,\bm{z}}] = 0$, for any $\bm{z}$, which requires $[X_{\bm{u}},E_{\alpha}] = 0$ for any $\alpha$. The last condition can only be satisfied by $X = \sum_\alpha x_{\alpha} E_{\alpha}$. Our specific measurement protocol is supported only on even bonds of the chain, while we know that $X_{\bm{u}}$ acts on both even and odd bonds, hence we can conclude that if $[X_{\bm{u}},h_{\rm{XX}}] = 0$ and $[X_{\bm{u}},\Gamma_{B,\bm{z}}] = 0$ then, necessarily, $X = 0$. 

For a generic conserved charge $Q_{\bm{u}}$, the conditional bath-charge distribution depends on the measurement outcome $\bm z$, as can be seen, for example, through its mean or variance. The measurement basis is thus partially revealing for a family of local conserved charges whose number grows linearly with the subsystem size $L_A$. This poses a question on the degree of universality of the projected ensemble, since we should track the effect of each outcome $\bm{z}$ on the subsystem. To discuss this, we closely follow the framework outlined in Ref.~\cite{chang-25}. We first identify a generator global state ensemble $\mathcal{F}$, which describes the characteristic late-time global states generated by the considered dynamics. For the $\rm{XX}$ dynamics studied here, a natural choice is to model $\mathcal{F}$ as a uniform distribution over all microstates compatible with the macrostate specified by an occupation function $n(k)$ fixed by the initial state.
Let us consider the case of infinite-temperature, in which $n(k) = 1/2$ for all $k$. We can model, in this case, the generator ensemble $\mathcal{F}_{\beta = 0}$ in a simpler way as
\begin{equation}
\label{eq:Fbeta0}
    \mathcal{F}_{\beta = 0} = \left\{\frac{1}{\binom{L}{L/2}}, \,\, \ket{\bm{n}} : \sum_{k=1}^L n_k = L/2 \right\}.
\end{equation}
Note that $\mathcal{F}_{\beta = 0}$ also contains states that are not microstates compatible with $n(k) = 1/2$. One example is $n_k = 1$ for $k = 1,\cdots,L/2$ and $n_k = 0$ otherwise. The fraction of these atypical states in $\mathcal{F}_{\beta = 0}$ is, however, vanishing in the thermodynamic limit. The reason behind this choice is the fact that the $\mathcal{F}_{\beta = 0}$-averaged state acquires the simple form
\begin{equation} \label{eq:Favgstate}
  \overline{\rho} = 
  \mathbb{E}_{\ket{\psi}\sim\mathcal{F}_{\beta = 0}}[\ket{\psi}\!\bra{\psi}] 
  = \frac{\Pi_{N,L/2}}{\binom{L}{L/2}},
\end{equation}
where we denoted by $\Pi_{O,\lambda}$ the projector onto the eigenspace of $O$ corresponding to the eigenvalue $\lambda$. The state in Eq.~\eqref{eq:Favgstate} can be decomposed into a sum of tensor products of projectors acting on $A$ and $B$ separately
\begin{equation} \label{eq:Favg}
  \overline{\rho} =\frac{1}{\binom{L}{L/2}} \sum_{n = 0}^{L_A} \Pi^{(A)}_{N_A,n} \otimes \Pi^{(B)}_{N_B,L/2-n}.
\end{equation}
Our goal is to understand the effect of an outcome $\bm{z}$ on average over the generator ensemble. The average state after the outcome $\ket{\bm{z}}$ is
\begin{equation}
  \rho_{\bm{z}} 
  = \text{Tr}_B\left[\,\overline{\rho} \,(\mathbb{I}_A \otimes \ket{\bm{z}}\!\bra{\bm{z}})
  \right]/p(\bm{z}),
\end{equation}
with $p(\bm{z}) = \text{Tr}\left[\,\overline{\rho}\,(\mathbb{I}_A \otimes \ket{\bm{z}}\!\bra{\bm{z}})\right]$ the associated Born probability. The conditional state $\rho_{\bm{z}}$ defines an equivalence relation between different outcomes, two outcomes $\bm{z}_1$ and $\bm{z}_2$ are equivalent $\bm{z}_1 \sim \bm{z}_2$ if $\rho_{\bm{z}_1} = \rho_{\bm{z}_2}$. Using the explicit form of $\overline{\rho}$ in Eq.~\eqref{eq:Favg}, we obtain
\begin{equation}
\label{eq:finalrhoz}
    \rho_{\bm{z}} = \frac{\sum_{n = 0}^{L_A} p_{\text{out}}(L/2-n|\bm{z}) \,\Pi^{(A)}_{N_A,n}}{\sum_{n = 0}^{L_A} p_{\text{out}}(L/2-n|\bm{z}) d^{(A)}_n},
\end{equation}
where $d^{(A)}_n = \binom{L_A}{n}$ is the dimension of the sector with particle number $n$ in $A$. 
We can make use of our knowledge of the post-measurement bath charge distribution $p_{\text{out}}(m|\bm{z}) = \bra{\bm{z}} \Pi^{(B)}_{N_B,m} \ket{\bm{z}}$ in the limit of large bath $L_B \to \infty$. From the distribution obtained in Eq.~\eqref{eq:pnbz}, we can assume $p_{\text{out}}(m|\bm{z})$ to approach a Gaussian form on the values compatible with the parity $\pi_{\bm{z}}$ fixed by $\bm{z}$
\begin{equation}
    p_{\text{out}}(m|\bm{z}) \simeq \frac{2}{\sqrt{2 \pi \sigma_{\bm{z}}^2}} \exp \left[-\frac{(m-\mu_{\bm{z}})^2}{2 \sigma_{\bm{z}}^2}\right],
\end{equation}
where $\mu_{\bm{z}} = L_B/2$ and, as we have discussed, for typical Born trajectories we have $\sigma_{\bm{z}}^2 = O(L_B)$ as $L_B \to \infty$. Since in Eq.~\eqref{eq:finalrhoz}, we are interested in $m = L/2-n$, we have $m-\mu_{\bm{z}} = L_A/2 -n$ which varies on a finite scale compared to the diverging bath size $L_B$. To compare the probabilities entering the conditional state~\eqref{eq:finalrhoz}, we therefore consider
\begin{equation}
\label{eq:ratiop}
    \frac{p_{\text{out}}(m + \delta |\bm{z})}{p_{\text{out}}(m |\bm{z})} \simeq \exp\left(-\frac{\delta^2}{2 \sigma_{\bm{z}}^2} - \frac{(m-\mu_{\bm{z}}) \delta}{\sigma_{\bm{z}}^2}\right),
\end{equation}
where $\delta = O(L_A)$ and even to respect the parity constraint. From Eq.~\eqref{eq:ratiop}, taking now the limit of large $L_B$ we obtain 
\begin{equation}
     \frac{p_{\text{out}}(m + \delta |\bm{z})}{p_{\text{out}}(m |\bm{z})} = 1 +O(L_A^2/L_B),
\end{equation}
plugging this result in Eq.~\eqref{eq:finalrhoz} we obtain the maximally mixed state in the subsystem with parity sector fixed by $\bm{z}$. Denoting the projector onto such space as $\Pi^{(A)}_{P_A,\pi_{\bm{z}}}$, we have
\begin{equation}
    \rho_{\bm{z}} = \frac{\Pi^{(A)}_{P_A,\pi_{\bm{z}}}}{2^{L_A-1}} + O\left(\frac{L_A^2}{L_B}\right),
\end{equation}  
which is, crucially, asymptotically independent of the outcome $\bm{z}$ within the fixed parity sector. Therefore, measurement outcomes are divided in only two equivalence classes, determined by the bath fermionic parity $\pi_{\bm{z}}$. To summarize, although most local charges are partially revealed by the measurement basis, in the thermodynamic limit, the information obtained from the measurements is completely washed out in $A$. Effectively, within this generator-ensemble description, all charges are maximally non-revealed to the subsystem as $L_B \to \infty$. As discussed in \cite{chang-25}, in this scenario, the pure states of $A$ in the projected ensemble are conjectured to be drawn from the uniform, maximally entropic manifold of states. In our case, this ensemble corresponds to the Gaussian Haar ensemble, defined as~\cite{bhk-21,bianchi-22,mcp-22}
\begin{equation}
\label{eq:GHEdef}
	\mathcal{E}_{\rm{GHE}} = \left\{ O \sim \text{Haar}(O(2 L_A)),\,\,\Gamma(O) = O J O^T\right\},
\end{equation}
in which $J = i \sigma_y \otimes \mathbb{I}_{L_A}$, the ensemble generates correlation matrices in the manifold $\Gamma \in O(2 L_A)/U(L_A)$~\cite{az-97}. This ensemble automatically takes into account the exact revelation of the fermionic parity discussed at the beginning of this section, since it is an equal mixture of the uniform ensemble on the two different parity components
\begin{equation}
\mathcal{E}_{\rm{GHE}} = p_{\rm{e}} \,\mathcal{E}_{\rm{GHEe}} +  p_{\rm{o}}\, \mathcal{E}_{\rm{GHEo}},
\end{equation}
with $p_{\rm{e}} = p_{\rm{o}} = 1/2$. The fixed parity ensemble $\mathcal{E}_{\rm{GHE}e}$ is obtained by taking ${O \sim \text{Haar}(SO(2L_A))}$, while $\mathcal{E}_{\rm{GHE}o}$ by multiplying each $O$ by some fixed reflection $R \in O(2L_A)$ with $\mathrm{det} R = -1$. The Gaussian Haar Ensemble $\mathcal{E}_{\rm{GHE}}$ and the fixed parity ensembles $\mathcal{E}_{\rm{GHE}e/o}$ produce exactly the same predictions when considering $\mathbb{E}[\Gamma^{\otimes k}]$ for $k<L_A$, while for $k\geq L_A$, the moments of the correlations matrix are generally different, making the difference in the two predictions observable, see App.~\ref{app:parityGHE} for a derivation. Moreover, as we show in App.~\ref{app:parityGHE}, the average entanglement entropy is unaffected by the presence of a single parity sector for all bipartitions. 

\subsection{Symmetry classification of the projected ensemble}
The Gaussian Haar ensemble $\mathcal{E}_{\rm{GHE}}$ in Eq.~\eqref{eq:GHEdef} uniformly samples the manifold of pure Gaussian states. The manifold of states explored by the projected ensemble, however, may be constrained by global symmetries preserved at the level of individual quantum trajectories. Here, we provide a general
classification of the trajectory symmetries and their constraints on the projected ensemble that can be realized by the unitary dynamics generated by $H_{\rm{XX}}$ combined with the measurement protocol in Eq.~\eqref{eq:Pialpha}.

In general, let $H$ be a quadratic Hamiltonian and $\{M_j\}_j$ be a set of mutually commuting measurement operators such that their measurement with outcome $\bm{z}$ provides a rank one projector on the bath $B$. We denote by $K_{\bm{z}} = \mathbb{I}_A \otimes \Pi_{\bm{z}}^{(B)}$ the Kraus operator associated with the non-unitary evolution due to the projection onto the measurement outcome. The evolution of an initial pure Gaussian state $\rho_0$ is
\begin{equation}
\check{\rho}_{\bm{z}} = K_{\bm{z}} e^{-i H t} \rho_0 e^{i H t} K^{\dagger}_{\bm{z}},
\end{equation}
here $\check{\rho}_{\bm{z}}$ is unnormalized and $p(\bm{z}) = \text{Tr}[\check{\rho}_{\bm{z}}]$ is the Born probability for the outcomes $\bm{z}$, while the normalized projected pure state in $A$ is $\rho_A(\bm{z}) = \text{Tr}_B[\check{\rho}_{\bm{z}}]/p(\bm{z})$. We say that the full system monitored dynamics has a symmetry if there exists a unitary (or antiunitary) operator $U$ such that, for any time $t$, we have 
\begin{equation}
\label{eq:traj_sym}
    U K_{\bm{z}} e^{- i H t} U^{-1}= K_{\bm{z}} e^{- i H t}, \qquad \text{ for any }\bm{z}.
\end{equation}
Equation \eqref{eq:traj_sym} implies, for $H$ and the measurement operator $M_j$,
\begin{equation}
\label{eq:sym}
    \begin{cases}
        U M_j U^{-1}= M_j\,\, \forall j, \qquad U H U^{-1} = H, \qquad &\text{if } U \text{ unitary},\\
        U M_j U^{-1}= M_j\,\, \forall j, \qquad U H U^{-1} = -H, \qquad &\text{if } U \text{antiunitary}.
    \end{cases}
\end{equation}
If we can find an operator $U$ satisfying these conditions and also the initial state is symmetric, i.e.\ $U \rho_0 U^{-1} = \rho_0$, the infinite temperature projected ensemble at late time is expected to be given, according to what we argued in the previous subsection, by the uniform measure over the manifold of $U-$invariant states.

Based on this logic, we now identify the Gaussian trajectory symmetries relevant to our dynamics and measurement protocol. Throughout this classification, we impose fermion-parity superselection and consider only transformations that act nontrivially within a fixed parity sector.
It is convenient to first analyze the condition on the measurement operators. This allows us to classify all of the stabilizers of $M_{j,\alpha}$ for $\alpha=1,2$ (see definition in Eq.~\eqref{eq:defM12}) and for a fixed $j$, which restricts us to the group $SO(4)$ of parity even (anti-)unitaries, acting on the Majorana fermions as $U_O \gamma_a U^{-1}_O = \sum_b O_{ab} \gamma_b$. We are labeling the Majorana on the physical sites $2j$ and $2j+1$ as $\gamma_a$ for $a = 1,\cdots,4$, such that $M_1 = i \gamma_{1} \gamma_3$ and $M_2 = i \gamma_2 \gamma_4$.

Requiring $U_O M_{\alpha} U^{-1}_O = M_{\alpha}$ further restricts the group of stabilizers into the subgroup $SO(2)_{13} \times SO(2)_{24}$ of independent rotations of the Majoranas present in each $M_{\alpha}$. It is now easy to explicitly verify that, if $U_O$ is unitary, a necessary (but not sufficient) condition for the Hamiltonian condition in Eq.~\eqref{eq:sym} being satisfied is the restriction of the rotation angles to either $0$ or $\pi$. This reduces the possibilities to a discrete subgroup $\mathbb{Z}_2$ of special orthogonal matrices such that $O_{a b} = \delta_{a b} \epsilon_a$, with $\epsilon_a = \pm 1$ and with $\epsilon_1 = \epsilon_3 = 1$ and $\epsilon_2 \epsilon_4 = 1$. Similarly, if $U_O$ is antiunitary, we only have a $\mathbb{Z}_2$ subgroup formed by $O_{a b} = \delta_{a b} \epsilon_a$ with $\epsilon_1 = - \epsilon_3 = -1$ and $\epsilon_2 \epsilon_4 = -1$.

Specialising to the case of the $\rm{XX}$ Hamiltonian, we find that, among the remaining valid operators, the only transformation satisfying \eqref{eq:traj_sym} is one of the two possible antiunitary transformations, we denote it by $\mathcal{T}$. Because of antiunitarity $\mathcal{T}i \mathcal{T}^{-1} = -i$, the action on the complex fermions operators is $\mathcal{T} c_j \mathcal{T}^{-1} = (-1)^{j+1} c_j$, while for Majorana operators
\begin{equation}
\label{eq:Tau_def}
\mathcal{T} \gamma_{j,1} \mathcal{T}^{-1} = (-1)^{j+1} \gamma_{j,1}, \qquad
\mathcal{T} \gamma_{j,2} \mathcal{T}^{-1} = (-1)^{j} \gamma_{j,2}.
\end{equation}
We have thus proven that there are only two possible candidates projected ensembles that can arise from our model. Then, according to the assumption of effective non-revelation of the conserved charges, if the initial state $\ket{\Psi_0}$ is not invariant under $\mathcal{T}$, i.e.\ $\mathcal{T}\ket{\Psi_0} \neq e^{i\varphi} \ket{\Psi_0}$ there is no symmetry constraint and the limiting projected ensemble explores the full manifold of Gaussian states uniformly, hence producing the ensemble $\mathcal{E}_{\rm{GHE}}$ in Eq.~\eqref{eq:GHEdef}.

If instead $\ket{\Psi_0}$ is invariant under $\mathcal{T}$, the manifold of explored states will be constrained such that each state in the ensemble remains $\mathcal{T}-$invariant. We denote the uniform distribution over such manifold as constrained Gaussian Haar Ensemble $\mathcal{E}_{\rm{cGHE}}$. After ordering the Majorana operators based on their $\mathcal{T}$ sign, we can write the Majorana correlation matrix as
\begin{equation}
    \Gamma = \begin{pmatrix}
        0 & R\\
        -R^T & 0
    \end{pmatrix},
\end{equation}
where $R \in O(L_A)$, sampling $R$ from the Haar measure in $O(L_A)$ defines the ensemble $\mathcal{E}_{\rm{cGHE}}$, see details in App.~\ref{app:cGHE}. In the remaining of this section we will consider both scenarios and verify numerically the emergence of these universal ensembles as late-time projected ensemble arising from Gaussian Hamiltonian dynamics.

\subsection{Numerical verifications}
In this subsection, we provide numerical evidence for the late-time convergence of the projected ensemble to our universal predictions. To do so, we numerically sample, according to the Born probability, the projected ensemble obtained from the state 
\begin{equation}
\ket{\Psi(t)} = e^{-i H_{\rm{XX}}t} \ket{\Psi_0}.
\end{equation}

To test the convergence of the projected ensemble to our universal predictions, we compute
\begin{equation}
  \Delta^{(k)} = \left \Vert \sum_{\bm{z}} p(\bm{z}) \Gamma^{\otimes k}(\bm{z}) 
  - \Gamma^{(k)}_{\mathcal{E}}\right \Vert_F,
\end{equation}
where $\left\Vert\cdot \right\Vert_F$ is the Frobenius norm, and $\mathcal{E}$ is the target universal ensemble. As a further check, following Ref.~\cite{lpdd-23}, we compute the projected ensemble average of the space-averaged von Neumann entanglement entropy of the subsystem $A$ defined as
\begin{equation} \label{eq:Savg}
\overline{S}_A(\bm{z}) = \frac{1}{L_A} \sum_{i = 1}^{L_A-1} S_{1}(\rho_{A_i} (\bm{z})),
\end{equation}
where $\rho_{A_i}(\bm{z}) = \text{Tr}_{\{i+1,\cdots,L_A\}}\left[ \ket{\psi(\bm{z})}\!\bra{\psi(\bm{z})}\right]$ and $S_1(\rho) = -\text{Tr}\left[\rho \log \rho\right]$ is the von Neumann entanglement entropy. We remark that the average over the projected ensemble $\overline{S}_A \equiv \mathbb{E}_{\bm{z} \sim \mathcal{E}_{\rm{PE}}} \left[\overline{S}_A(\bm{z})\right]$ depends on all the moments of $\mathcal{E}_{\rm{PE}}$, thus providing a non-trivial check on the full hierarchy of moments. 

\begin{figure*}[t]
\centering
\includegraphics[width=0.99\textwidth]{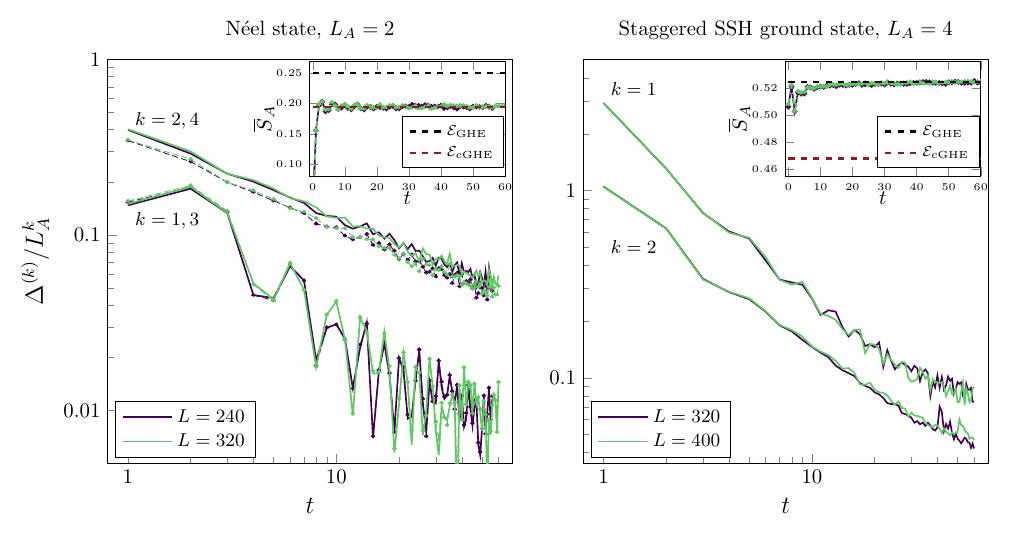}
    \caption{\label{fig:infT} Left panel: rescaled Frobenius distance $\Delta^{(k)}/L_A^k$ between the correlation matrix moments of the numerically sampled projected ensemble in Eq.~\eqref{eq:PEmom} obtained from the infinite temperature Néel quench and the moments of the Gaussian Haar ensemble constrained by the antiunitary symmetry $\mathcal{T}$ in Eq.~\eqref{eq:Tau_def} for a subsystem with $L_A = 2$ sites. Solid lines correspond to $k = 1,2$ and dashed lines to $k = 3,4$. The distances for different moments $k$  approach zero with a power law behaviour indicating convergence of the projected ensemble to the universal constrained Haar Ensemble $\mathcal{E}_{\rm{cGHE}}$. In the inset we show the time evolution of the space-averaged $\overline{S}_A$ von Neumann entanglement entropy of the subsystem $A$ as in Eq.~\eqref{eq:Savg} averaged over the projected ensemble. The entropy quickly saturates to a value compatible with the prediction of $\mathcal{E}_{\rm{cGHE}}$ (red dashed line). Right panel: Same as left for initial state given by the ground state of the staggered SSH model such that the $\mathcal{T}$-symmetry is broken. In this case we consider a subsystem of $L_A = 4$ sites and compute the rescaled distance $\Delta^{(k)}/L_A^k$ with respect to the Gaussian Haar Ensemble moments. The inset shows the saturation of $\overline{S}_A$ to the value predicted by $\mathcal{E}_{\rm{GHE}}$ (black dashed line).
    In all datasets, for each time we reconstruct the moments from $\sim 10^4$ samples.
    }
\end{figure*}

We start our numerical analysis by considering, as initial state, the Néel state 
\begin{equation}
\ket{\rm{N}} = \left(\prod_{k=1}^{L/2} c^{\dagger}_{2k}\right) \ket{0},
\end{equation}
with $\ket{0}$ the vacuum of the complex fermions $c_j,c^{\dagger}_j$. We observe that $\mathcal{T} \ket{\rm{N}} = - \ket{\rm{N}}$, hence we predict the convergence of $\mathcal{E}_{\rm{PE}}$ to the constrained ensemble $\mathcal{E}_{\rm{cGHE}}$. The left panel of Fig.~\ref{fig:infT} numerically confirms our prediction.
The distances $\Delta^{(k)}$ for $k = 1,2,3,4$ go to zero exhibiting a power law behaviour as already observed for local density measurements in \cite{lpdd-23}. The space-averaged entanglement entropy is also in agreement with the prediction from the ensemble $\mathcal{E}_{\rm{cGHE}}$ we have proposed based on the analysis of the discrete symmetries constraining the explored manifold of projected states and on the assumption of the effective non-revelation of all conserved charges. 

As an initial state breaking the $\mathcal{T}-$ symmetry locally relaxing to an infinite temperature state we can choose the ground state of the critical staggered Su–Schrieffer–Heeger (SSH) chain 
\begin{equation}
H_{\rm{SSH}} = \sum_j (-1)^j c^{\dagger}_j c_{j+1} + h.c.,
\end{equation}
given by 
\begin{equation}
\ket{\Psi_{\rm{SSH}}} = \prod_{|k| < \pi/2} \frac{c^{\dagger}_k + i\, \text{sgn}(\sin k)\,c^{\dagger}_{k+\pi}}{\sqrt{2}} \ket{0}.
\end{equation}
Explicit computations show: (i) $\langle\Psi_{\rm{SSH}}|\hat n_k|\Psi_{\rm{SSH}}\rangle=1/2$ for every $k$, and (ii) $\mathcal{T} \ket{\Psi_{\rm{SSH}}} \neq e^{i \varphi} \ket{\Psi_{\rm{SSH}}}$.

In the right panel of Fig.~\ref{fig:infT} we verify the convergence of the late-time projected ensemble to the unconstrained Gaussian Haar random ensemble $\mathcal{E}_{\rm{GHE}}$ (see Eq.~\eqref{eq:GHEdef}) from this initial state.
In Appendix~\ref{app:Ctridiag} we show how to construct a family of states satisfying (i) and (ii).

\section{Finite-temperature projected ensemble}
\label{sec:finittemp}
In this section, we consider the case of a quench locally relaxing to a finite temperature GGE. We use as initial state a dimer state
\begin{equation}
\label{eq:dimer}
 \ket{\Psi_0(\alpha)} = \prod_{j=1}^{L/2} \frac{c^{\dagger}_{2j-1} + \alpha c^{\dagger}_{2j}}{\sqrt{1 + |\alpha|^2}} \ket{0}.
\end{equation}
This state, time evolved with the Hamiltonian~\eqref{eq:Hxx}, locally thermalizes to a GGE $\rho_{\rm{GGE}}(\alpha)$ such that $\text{Tr}[\rho_{\rm{GGE}}(\alpha) n_k] = \bra{\Psi_0(\alpha)}n_k \ket{\Psi_0(\alpha)} \equiv n(k)$. In particular, for the dimer state in Eq.~\eqref{eq:dimer}, one has
\begin{equation}
\label{eq:nk_dimer}
n(k) = \frac{1}{2} + \text{Re}\left[\frac{e^{-ik} \alpha}{1+ |\alpha|^2}\right].
\end{equation}

A first attempt to characterize the finite-temperature projected ensemble for the same free-fermionic unitary dynamics studied in this work but with local density measurements was proposed in Ref.~\cite{lpdd-23}, where the authors define a representative state $\ket{\Phi}$ of the global system whose conserved charges densities coincide, in the thermodynamic limit, with those of the initial state $\ket{\Psi_0(\alpha)}$. The representative state $\ket{\Phi}$ in the bath is then projectively measured, with outcomes sampled from the corresponding Born probability. The resulting ensemble of pure states in $A$ obtained in this way was termed \textit{deep}-GGE and in the following will be denoted as $\mathcal{E}_{\rm{dGGE}}$. It was shown that $\mathcal{E}_{\rm{dGGE}}$ numerically reproduces the finite moments of the projected ensemble obtained from the unitary dynamics when approaching the limit $L_B \to \infty$ and, subsequently, $t\to \infty$. As an explicit example, a simple possible way to fix the representative state $\ket{\Phi}$ for the $\rm{XX}$ dynamics consists in choosing a single eigenstate of $H_{\rm{XX}}$ at half-filling 
\begin{equation}
\ket{\Phi} = c^{\dagger}_{k_1}\cdots c^{\dagger}_{k_{L/2}} \ket{0}.
\end{equation}
The occupied modes are chosen such that the density of occupied modes reproduces, in the limit $L\to\infty$, the density $n(k)$ of occupied modes of the initial dimer state (Eq.~\eqref{eq:nk_dimer}). In practice, one asks that in an interval $[k, k + \Delta k]$ there are $L/(2\pi) \, n(k) \Delta k$ occupied modes.

Although the deep-GGE construction numerically reproduces the limiting projected ensemble, it is formulated in terms of a \textit{global} representative state and requires the explicit evaluation of the Born probabilities associated with the bath measurements. In the following we solve these problems by introducing a Scrooge ensemble on the manifold of fermionic Gaussian states, defined only in the \emph{local} subsystem of interest, which we argue describes the limiting projected ensemble. Interestingly, a similar constrained analogue of the Scrooge ensemble was obtained in continuous variable systems~\cite{lhh-24}, here we argue for its emergence in a deterministic non-interacting fermionic quantum system.

\subsection{Fermionic Gaussian Scrooge Ensemble}

Let $\ket{\phi_{\Gamma}}$ be a pure Gaussian state with Majorana correlation matrix $\Gamma$, and let $\rho_{\rm{GGE}}$ be the Generalized Gibbs ensemble attained by the unitary dynamics in the limit $L_B \to \infty$ and $t\to \infty$. We then define the Gaussian Scrooge Ensemble as the following ensemble of pure states in subsystem $A$\begin{equation}
\label{eq:GScrooge}
    \mathcal{E}_{GS} = \left\{ D_A \bra{\phi_{\Gamma}} \rho_{\mathrm{GGE},A} \ket{\phi_{\Gamma}} d\mu_{\rm{H}}(\Gamma) , \,\,\,\ket{\psi_{\Gamma'}} = \frac{\sqrt{\rho_{\mathrm{GGE},A}} \ket{\phi_{\Gamma}}}{\sqrt{\bra{\phi_{\Gamma}} \rho_{\mathrm{GGE},A}\ket{\phi_{\Gamma}}}}\right\},
\end{equation}
where $d\mu_{\rm{H}}(\Gamma)$ denotes the invariant measure over the manifold $O(2L_A)/U(L_A)$ of Gaussian pure states and $D_A = 2^{L_{A}}$ is the dimension of the Hilbert space of $A$. It is easy to check that the first moment of the Gaussian Scrooge Ensemble coincides with the subsystem reduced $\rho_{\mathrm{GGE},A}$
\begin{equation}
%\begin{split}
\mathbb{E}_{\ket{\psi}\sim\mathcal{E}_{GS}}[ \ket{\psi}\!\bra{\psi}] = 
  D_A \sqrt{\rho_{\mathrm{GGE},A}} \left( \int d\mu_{\rm{H}}(\Gamma) 
  \ket{\phi_{\Gamma}}\!\bra{\phi_{\Gamma}}\right) \sqrt{\rho_{\mathrm{GGE},A}} 
  =\rho_{\mathrm{GGE},A},
%\end{split}
\end{equation}
where we used the fact that $\int d\mu_{\rm{H}}(\Gamma)\ket{\phi_{\Gamma}}\!\bra{\phi_{\Gamma}} = \mathbb{I}/D_A$. The effect of the $\sqrt{\rho_{\mathrm{GGE},A}}$-distortion of the Gaussian Haar Ensemble can be encoded in the Majorana correlation matrix as the following non-linear transformation
\begin{equation}
\label{eq:gammadist}
    \Gamma' = \Gamma_{\mathrm{GGE},A} + S \Gamma (\mathbb{I}-\Gamma_{\mathrm{GGE},A} \Gamma)^{-1} S,
\end{equation}
where $ S = \sqrt{\mathbb{I} + \Gamma_{\mathrm{GGE},A}^2}$, see App.~\ref{app:rhodist} for the derivation. The reweighting factor of the Haar measure is explicitly given by
\begin{equation}
    \bra{\phi_{\Gamma}} \rho_{\mathrm{GGE},A}\ket{\phi_{\Gamma}} = 2^{-L_A} \sqrt{\det(\mathbb{I} -\Gamma_{\mathrm{GGE},A}\Gamma)}.
\end{equation}

To argue for the emergence of $\mathcal{E}_{GS}$ we should, as before, show that charges are not revealed by the measurements. In this case, however, we cannot use our previous argument based on the global generator ensemble. Indeed, a simple decomposition of $\mathcal{F}$-averaged state as the one in Eq.~\eqref{eq:Favg} is not available at finite temperature. Nevertheless, similar to the post-measurement particle number distribution in Eq.~\eqref{eq:pnbz}, in App.~\ref{app:variance} we show that generic conserved charges display an extensive variance for Born-typical trajectories. This suggests that, in the thermodynamic limit, the effective non-revelation persists. 

\begin{figure*}[t]
\centering
\includegraphics[width=0.99\textwidth]{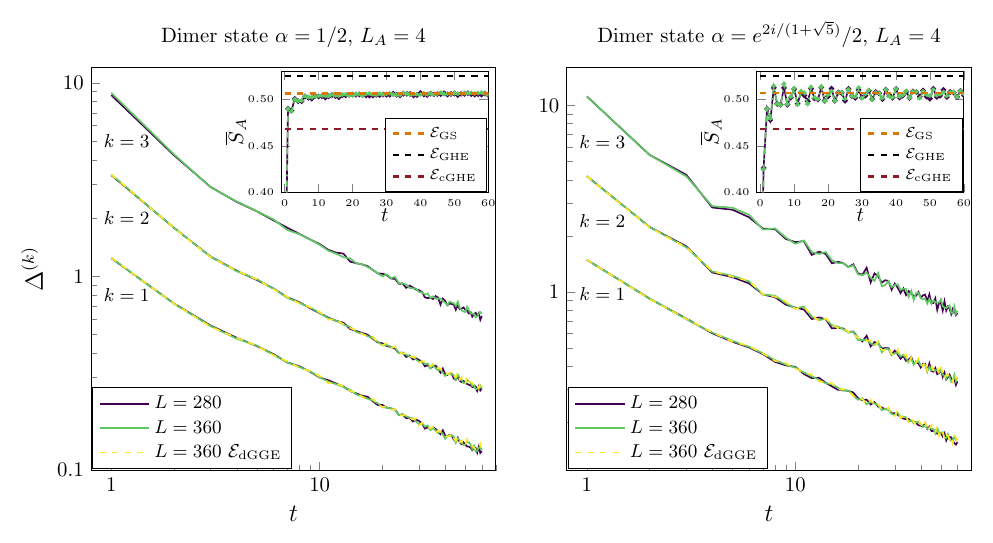}
    \caption{\label{fig:finT} Left panel: Frobenius distance $\Delta^{(k)}$ between the correlation matrix moments of the numerically sampled projected ensemble in Eq.~\eqref{eq:PEmom} obtained from the finite temperature dimer quench with $\alpha = 0.5$ and the moments obtained from the Gaussian Scrooge ensemble (solid lines) on a subsystem of $L_A = 4$ sites, defined in Eq.~\eqref{eq:GScrooge}. The dashed line shows the same quantity but when computing the distance $\Delta^{(k)}$ with respect to the \textit{deep}-GGE obtained by performing the measurements on a global representative state~\cite{lpdd-23}, the results are indistinguishable within the statistical uncertainty. Inset: Time evolution of the space-averaged $\overline{S}_A$ von Neumann entanglement entropy of the subsystem $A$ as in Eq.~\eqref{eq:Savg}. The entropy quickly saturates to a value compatible with the prediction of the fermionic Gaussian Scrooge ensemble $\mathcal{E}_{\rm{GS}}$ (orange dashed line). The moments of the projected ensemble at a fixed time are obtained from $\sim 10^4$ samples. Right panel: Same as left but for a different dimer initial state with $\alpha = e^{2i/(1+\sqrt{5})}/2$.}
\end{figure*}

We verify this prediction numerically, by determining exactly the reduced correlation matrix $\Gamma_{\mathrm{GGE},A}$ for the initial state $\ket{\Psi_0(\alpha)}$ (Eq.~\eqref{eq:dimer}) and then sampling numerically the fermionic Gaussian Scrooge ensemble in Eq.~\eqref{eq:GScrooge}.
Note that the initial dimer state $\ket{\Psi_0(\alpha)}$ is not $\mathcal{T}$-invariant for $\mathfrak{Re}\, \alpha \neq 0$, so no constraint is applied to the manifold of Gaussian states explored by the Haar measure $d\mu_{\rm{H}}(\Gamma)$. In Fig.~\ref{fig:finT}, we compare the finite-time projected ensemble with the predictions of the Gaussian Scrooge ensemble and the single-eigenstate approach of Ref.~\cite{lpdd-23}. The results agree within statistical uncertainties, supporting the Gaussian Scrooge ensemble as a description of the universal limiting projected ensemble formulated directly at the subsystem level.

\section{Conclusions}
\label{sec:concl}

In this manuscript, we have investigated deep thermalization in the $\rm{XX}$ chain under local projective measurements preserving Gaussianity while breaking particle number conservation. We have provided numerical evidence that the limiting projected ensemble at finite temperature of Gaussian Hamiltonian dynamics is given by a fermionic Gaussian Scrooge ensemble, obtained as a distortion of the Haar measure over Gaussian states such that the first moment matches the Generalized Gibbs ensemble determined by the initial state of the quench. This result provides a universal description of the projected ensemble in an integrable setting with an extensive family of conservation laws directly formulated on the subsystem, extending prior results based on full-system properties~\cite{lpdd-23}. We have also identified how symmetries preserved along individual measurement trajectories restrict the accessible Gaussian-state manifold and distinguish the corresponding universal ensembles.

The physical mechanism underlying this description is the effective non-revelation of local conserved charges. Although individual measurement outcomes retain information about bath charges at finite size, this information becomes irrelevant to a finite subsystem as the bath becomes larger. This suggests that a measurement basis need not be maximally charge non-revealing at finite size to produce a Scrooge ensemble. Our analysis thus applies the charge-revelation framework of Ref.~\cite{chang-25} to integrable Gaussian dynamics, identifying conditions under which partially revealing measurements become effectively non-revealing for a large bath.

Several questions remain open. A natural direction is to extend our analysis to interacting integrable models and investigate whether effective non-revelation leads to a Scrooge ensemble determined by the reduced GGE, or whether additional signatures of integrability persist in the higher moments of the projected ensemble.

Gaussian dynamics could also provide insights into the role of conservation laws in mixed-state deep thermalization, arising from either incomplete bath measurements~\cite{yhk-25} or a mixed initial state~\cite{sr-26}.

Finally, exploring the relation between deep thermalization and measurement-induced criticality~\cite{srn-19,abd-21,poboiko-23,fava-23} offers a promising direction~\cite{ih-23,wu-26}. Building on previous studies on the role of symmetries~\cite{vasseur-22,fava-24} and their breaking~\cite{rac-26} in monitored quantum systems, our results could help clarify how these features shape the universal properties of measurement-conditioned ensembles in Gaussian fermionic systems.
\\

{\bf Note}.--- After the completion of this manuscript, but prior to its submission, a preprint appeared on arXiv introducing the same definition of the fermionic Gaussian Scrooge ensemble \cite{sz-26}. Apart from this common definition, however, the content and scope of the two works are different.

\medskip

\textit{Acknowledgments.---} We thank Riccardo Senese for useful discussions. AR and PC  acknowledge support from the European Research Council under the Advanced Grant No.\ 101199196 (MOSE). BB acknowledges financial support from the Royal Society through the University Research Fellowship No.\ 201101 and from the European Research
Council under the Consolidator Grant No.\ 101231030 (DREAM).  Views and opinions expressed are however those of the author(s) only and do not necessarily reflect those of the European Union or the European Research Council. Neither the European Union nor the granting authority can be held responsible for them.

Use of generative AI statement: We used ChatGPT 5.6 for minor text polishing and code optimization.

\appendix

\section{Correlation matrix after projective measurements of Majorana bilinears}
\label{app:app_Gammaz}

In this appendix, we derive the transformation law of the Majorana covariance matrix under a projective measurement of a Majorana bilinear.

Consider a Gaussian state $\ket{\Psi}$ defined on a system of $L$ qubits. By definition, the state is fully characterized by the covariance matrix of Majorana fermions
\begin{equation}
\Gamma_{ab} = \frac{i}{2} \bra{\Psi} [\gamma_a,\gamma_b] \ket{\Psi} = i (\bra{\Psi} \gamma_a \gamma_b \ket{\Psi} - \delta_{a,b}),
\end{equation}
where $a,b = 1,\cdots, 2 L$. We order the Majorana operators as $\gamma_{2k-1} \equiv \gamma_{k,1} $ and $\gamma_{2k} \equiv \gamma_{k,2}$ for $k = 1,\cdots,L$. The covariance matrix is real and antisymmetric.

Let us consider the general problem of the evolution of the correlation matrix after the projective measurement of the Gaussianity preserving operator
\begin{equation}
\mathcal{O}_{cd} = i \gamma_c \gamma_d,
\end{equation}
 for generic $c\neq d$. The case of interest of the operator $M_1$ ($M_2$)  in Eq.~\eqref{eq:defM12} acting on the sites $2j-1$ and $2j$ is recovered for $\mathcal{O}_{4j-3,4j-1}$ ($\mathcal{O}_{4j-2,4j}$).
For a generic outcome $z = \pm 1$, the projector is
\begin{equation}
\Pi_z = \frac{\mathbb{I}+z \mathcal{O}_{cd}}{2} = \frac{\mathbb{I} +i z \gamma_{c} \gamma_{d} }{2}.
\end{equation}
Therefore, the post-measurement correlation matrix is given by, for $a \neq b$, 
\begin{equation}
\label{eq:Gammaz}
\Gamma_{ab}(z) = \frac{i}{2} \frac{\bra{\Psi} \Pi_z [\gamma_a,\gamma_b] \Pi_z  \ket{\Psi}}{\bra{\Psi} \Pi_z  \ket{\Psi} } = i \frac{\bra{\Psi} \Pi_z \gamma_a \gamma_b \Pi_z  \ket{\Psi}}{\bra{\Psi} \Pi_z  \ket{\Psi} }.
\end{equation}
The denominator on the right-most side of Eq.~\eqref{eq:Gammaz} is the Born probability $p(z) = \bra{\Psi} \Pi_z \ket{\Psi}$ for the outcome $z$ and it is given by
\begin{equation}
p(z) = \bra{\Psi} \Pi_z  \ket{\Psi}  = \frac{1}{2} (1 + z \Gamma_{cd}).
\end{equation}
Writing explicitly \eqref{eq:Gammaz}, we obtain
\begin{equation}
\label{eq:Gammaz_1}
\Gamma_{ab}(z) = \frac{2i}{1+ z \Gamma_{cd}} \frac{1}{4}\left( \langle \gamma_a\gamma_b\rangle + z \langle \mathcal{O}_{cd} \gamma_a \gamma_b \rangle + z \langle \gamma_a \gamma_b  \mathcal{O}_{cd} \rangle + \langle  \mathcal{O}_{cd} \gamma_a \gamma_b  \mathcal{O}_{cd} \rangle \right)
\end{equation}
We distinguish three different cases in the computation of \eqref{eq:Gammaz_1}:

\textbf{I.} If $a = c$ and $b = d$ we can use $\mathcal{O}_{cd} \gamma_c \gamma_d =  \gamma_c \gamma_d  \mathcal{O}_{cd} = i  \gamma_c \gamma_d \gamma_c \gamma_d   = -i \mathbb{I}$
\begin{align}
\label{eq:1}
\Gamma_{cd}(z) &= \frac{2i}{1+ z \Gamma_{cd}} \frac{1}{2}\left( \langle \gamma_c \gamma_d\rangle -  i z \right)  \\
&= \frac{1}{1+ z \Gamma_{cd}} (\Gamma_{cd} +z ) \\
& = z.
\end{align}
 
\textbf{II.} If $a = c \text{ or } d$ and $b \neq c,d$ we have $\mathcal{O}_{cd} \gamma_c \gamma_b = -i \gamma_d \gamma_b $ and also $\mathcal{O}_{cd} \gamma_c \gamma_b \mathcal{O}_{cd}  = - \gamma_c \gamma_b$  hence 
\begin{equation}
\label{eq:2}
\Gamma_{cb}(z) = 0,
\end{equation}
same holds for $\Gamma_{db}(z) = \Gamma_{ac}(z) = \Gamma_{ad}(z) = 0$.

\textbf{III.} If $a,b \neq c,d$ we can use the fact that $[\mathcal{O}_{cd}, \gamma_a \gamma_b ] = 0$ and $\mathcal{O}_{cd}^2 = \mathbb{I}$ and write
\begin{align}
\Gamma_{ab}(z) &=  \frac{i}{1+ z \Gamma_{cd}}  (\langle \gamma_a\gamma_b\rangle + i z \langle\gamma_a \gamma_b \gamma_c \gamma_d \rangle), 
\end{align}

using Wick's theorem we obtain
\begin{align}
\label{eq:3}
\Gamma_{ab}(z) &= \frac{1}{1+z \Gamma_{cd}} ( \Gamma_{ab} + z (\Gamma_{ab}\Gamma_{cd} - \Gamma_{ac}\Gamma_{bd} + \Gamma_{ad}\Gamma_{bc}))  \\
&= \Gamma_{ab} + \frac{z}{1+ z \Gamma_{cd}} \left(\Gamma_{ad}\Gamma_{bc}- \Gamma_{ac}\Gamma_{bd}   \right).
\end{align}
The equations \eqref{eq:1}-\eqref{eq:2}-\eqref{eq:3} can be written more compactly. For a generic correlation matrix 
\begin{equation}
\Gamma = 
\begin{pmatrix}
\Gamma' & S \\
-S^T &
D_{cd}
\end{pmatrix},
\end{equation}
with $D_{cd} = \begin{pmatrix}
    0 & \Gamma_{cd}\\
    -\Gamma_{cd} & 0 
\end{pmatrix}$ the $2\times2$ block before the measurement, the post-measurement $\Gamma(z)$ is 
\begin{equation}
\label{eq:measbilin}
    \Gamma(z) = \begin{pmatrix}
        \Gamma' + S(D_{cd}+ M_z)^{-1}S^{T} & 0 \\
        0 & M_z
    \end{pmatrix},
\end{equation}
where $M_z \equiv \begin{pmatrix}
    0 & z \\
    -z & 0
\end{pmatrix}$. 
Equation \eqref{eq:measbilin} can be generalized to simultaneous measurement of disjoint Majorana bilinears \cite{bravyi-05}. Consider the measurement on the full bath $B$, we have an initial $\Gamma$ given by 
\begin{equation}
    \Gamma = \begin{pmatrix}
        \Gamma_A & \Gamma_{AB} \\
        -\Gamma_{AB}^{T} & \Gamma_B
    \end{pmatrix},
\end{equation}
the post-measurement correlation matrix $\Gamma_A(\bm{z})$ in the subsystem $A$ is
\begin{equation}
\label{eq:proj_formula}
\Gamma_A(\bm{z}) = \Gamma_A + \Gamma_{AB}(\Gamma_B + \Gamma_{B,\bm{z}})^{-1} \Gamma_{AB}^T,
\end{equation}
where $\Gamma_{B,\bm{z}} = \bigoplus_{j = 1}^{L_B} \begin{pmatrix}
    0 & z_j\\
    -z_j & 0
\end{pmatrix}$ is the correlation matrix of the post-measurement state in $B$.

\section{Gaussian Haar Ensembles}
\label{app:Haar}

In this appendix, we describe some details on the Gaussian Haar ensemble used in the main text. Subsection~\ref{app:parityGHE} discusses the role of having a parity constraint in the Gaussian Haar ensemble~\eqref{eq:GHEdef}. Subsection~\ref{app:cGHE} provides a construction of the Gaussian Haar ensemble in the presence of the antiunitary $\mathcal{T}$ ~\eqref{eq:Tau_def}.

\subsection{Effects of parity constraint on $\mathcal{E}_{\rm{GHE}}$}
\label{app:parityGHE}

Here, we show that the correlation matrix moments $\mathbb{E}[\Gamma^{\otimes k}]$ of the equal mixture of parity sectors and the fixed-parity Haar ensemble differs for $k\geq L$ where $2L$ is the number of Majorana fermions.

To show this, note that, because of the left/right invariance of the Haar measure, all the moments must satisfy $\mathbb{E}_O[\Gamma^{\otimes k}] = R^{\otimes 2 k}\, \mathbb{E}_O[\Gamma^{\otimes k}] $ for any $R$ belonging to the group on which the Haar measure is defined. As a consequence, the moments $\mathbb{E}_O[\Gamma^{\otimes k}]$ must be written in terms of the invariant tensors of the group under consideration. We now compute the moments over the group $O(2L)$ and over the connected component $SO_{\pm}(2 L)$ with fixed parity $\det O =  1$ and the coset $SO_-(2L)$ obtained by multiplying a reflection to any element of $SO_+(2 L)$. For moment order $k < L$, the only invariant tensors are given by combinations of Kronecker deltas for all the groups, which implies that the two different ensemble give exactly the same moments. For index $k\geq L$ an additional invariant tensor is present only for $SO(2L)$, the tensor is given by the Levi-Civita tensor $\epsilon_{i_1\cdots i_{2L}}$ and the invariance property can be explicitly derived from the property
\begin{equation}
R_{i_1 j_1} \cdots R_{i_{2L} j_{2L}} \epsilon_{j_1 \cdots j_{2L}} = \det R\, \epsilon_{i_1 \cdots i_{2L}} = \epsilon_{i_1 \cdots i_{2L}}
\end{equation}
where $\det R = 1$ since $R\in SO(2L)$. For $R\in O(2 L)$, $\epsilon$ is therefore not invariant. Let us define
\begin{equation}
    T^{(k)} \equiv \mathbb{E}_{O \sim \text{Haar}(O(2L))}[\Gamma^{\otimes k}],
\end{equation}
which only contains invariant tensor coming from Kronecker delta combinations. Then, for $SO(2L)$ and the coset $SO_-(2 L)$ we must also include the Levi-Civita contribution. For $k=L$ there is only one additional term
\begin{equation}
    T_{\pm}^{(L)} = T^{(L)} + \alpha_{\pm} \,\epsilon.
\end{equation}
To determine $\alpha_{\pm}$ we use the property $\epsilon_{i_1 \cdots i_{2 L}} \epsilon_{i_1 \cdots i_{2 L}} = (2 L)!$ and the fact that the contraction of $\epsilon$ with $T^{(L)}$ is identically zero, because $T^{(L)}$ only contains Kronecker deltas. Hence, we have
\begin{equation}
    \epsilon\cdot T^{(L)}_{\pm} = \alpha_{\pm} (2 L)!.
\end{equation}
The contraction in the left hand side is computed as $\epsilon \cdot T_{\pm}^{(L)}  = 2^{L} L! \,\text{Pf}(\Gamma) = 2^L L! \det O \,\text{Pf}(J) = \pm2^{L} L!$, where we have used the definition of the Pfaffian~\cite{bravyi-05}. Finally we have 
\begin{equation}
\label{eq:paritymom}
    T^{(L)}_{\pm} = T^{(L)} \pm \frac{1}{(2L-1)!!} \epsilon.
\end{equation}
The same logic applies to higher moments $k >L$, but the explicit derivation becomes cumbersome because of the many possible invariant tensors that can be constructed by multiplying Kronecker deltas and the Levi-Civita tensor. To summarize, Eq.~\eqref{eq:paritymom} explicitly shows the differences between the moments of the Gaussian Haar Ensemble with fixed parity and the equal mixture. These differences can thus be observed numerically.

We now analyze the effect on the entanglement entropy of a bipartition of the pure state in the two ensembles. We can prove that the entanglement spectrum is exactly unaffected by the presence of a single parity sector. To prove this we need to show that, given a subsystem $A$ identified by $2 L_A$ Majorana operators, the Majorana correlation matrix restricted to $A$ has a spectrum that does not depend on the parity sector. In particular, let $O$ be distributed according to the Haar measure over $SO(2 L)$. To sample over the odd-parity sector we just need to multiply $O$ by a reflection $R$ such that $R^TR=\mathbb{I}$ and $\det R = -1$. The correlation matrix then transforms as $\Gamma \to R\Gamma R^T$. Using the left/right invariance of the Haar measure, we can always choose $R$ to be of the form $R = \mathbb{I}_A \oplus R_B$. It is now easy to show that, with this choice, the restriction of $\Gamma$ to $A$ is invariant and therefore, the spectrum is left unchanged.

\subsection{$\mathcal{T}-$constrained Gaussian Haar Ensemble $\mathcal{E}_{\rm{cGHE}}$}
\label{app:cGHE}

Here we construct the Gaussian Haar ensemble of pure states constrained to be invariant under the antiunitary operator $\mathcal{T}$ defined in Eq.~\eqref{eq:Tau_def}. Consider an initial ordering of the Majorana fermions $\{\gamma_a\}_{a = 1,\cdots,2 L}$, we can always reorder them with a special orthogonal rotation $R \in SO(2L)$ such that, in the new basis, the single-particle action of $\mathcal{T}$, i.e. $\mathcal{T} \bm{\gamma} \mathcal{T}^{-1} \equiv T \bm{\gamma}$, has the form 
\begin{equation}
\label{eq:appTmaj}
    T = \begin{pmatrix} \mathbb{I}_{L}  & 0 \\
0 &- \mathbb{I}_{L} \end{pmatrix}.
\end{equation}
For example, this would be achieved by the ordering $\{\gamma_{1,1}, \gamma_{2,2}, \cdots ,\gamma_{L,2}, \gamma_{1,2} ,\gamma_{2,1}, \cdots, \gamma_{L,1}\}$. In this basis, the correlation matrix has the form 
\begin{equation}
\label{eq:block}
\Gamma = \begin{pmatrix} 0 & X \\
-X^{T} & 0 
\end{pmatrix},
\end{equation}
where $X \in O(L)$. In any basis in which Eq.~\eqref{eq:appTmaj} is valid, the constrained Gaussian Haar Ensemble is defined as
\begin{equation}
\label{eq:cGHEdef}
\mathcal{E}_{\rm{cGHE}} = \left\{ X \sim \text{Haar}(O(L)),\, \Gamma(O) =  OJO^T \text{ with } O = \begin{pmatrix} X & 0 \\
0 & \mathbb{I}
 \end{pmatrix}\right\},
\end{equation}
where again $J = i \sigma_y \otimes \mathbb{I}_{L}$. As one can show explicitly, the $U(L)$ redundancy present in $\mathcal{E}_{\rm{GHE}}$ \eqref{eq:GHEdef} disappears once we impose the block structure for $O$ in the definition \eqref{eq:cGHEdef}. Thus, the ensemble $\mathcal{E}_{\rm{cGHE}}$ describes a uniform probability distribution of pure Gaussian states over an $O(L)$ manifold.

The first few moments of $\mathcal{E}_{\rm{cGHE}}$ can be easily computed analytically. The first moment vanishes

\begin{equation}
\Gamma^{(1)}_{\rm{cGHE}} = \mathbb{E}_{\Gamma \sim \mathcal{E}_{\rm{cGHE}}} \left[ \Gamma \right] = 0,
\end{equation}
where we used $\mathbb{E}_{X \sim \text{Haar}(O(L))}[X] = 0$. The second moment is given by
\begin{equation}
\Gamma^{(2)}_{\rm{cGHE}} = \mathbb{E}_{\Gamma \sim \mathcal{E}_{\rm{cGHE}}} \left[ \Gamma^{\otimes 2} \right] = W^{(2)},
\end{equation}
the coordinates of the tensor $W^{(2)}$ are
\begin{equation}
W^{(2)}_{(i,\mu) (j,\nu) , (k,\rho) (l,\sigma)} = \frac{1}{L} \times\begin{cases} \delta_{ik} \delta_{jl} &\quad \mu <\nu, \rho<\sigma \\ 
-\delta_{il} \delta_{jk} &\quad \mu <\nu, \rho>\sigma \\
-\delta_{jk} \delta_{il} &\quad \mu >\nu, \rho<\sigma \\
\delta_{ik} \delta_{jl} &\quad \mu >\nu, \rho>\sigma
\end{cases},
\end{equation}
and $0$ otherwise. The Latin indices go from $1$ to $L$ while the Greek indices are $0,1$ and label the two eigenspaces of $T$. Higher moments can be computed in terms of standard averages over the orthogonal group~\cite{cs-06}. 

\section{$\mathcal{T}-$noninvariant Gaussian infinite temperature initial state}
\label{app:Ctridiag}
In this appendix, we describe how to construct pure Gaussian initial states
with fixed particle number that break the antiunitary
symmetry $\mathcal T$ defined in Eq.~\eqref{eq:Tau_def}
and satisfy
$\langle\Psi_0|\hat n_k|\Psi_0\rangle=1/2$ for every $k$.
Consider, for the complex fermions correlation matrix $C_{ij} = \bra{\Psi} c^{\dagger}_i c_j\ket{\Psi}$, a generic tridiagonal Hermitian matrix 
\begin{equation}
    C_{ij} = \frac{1}{2} \delta_{ij} + a_j \delta_{i,j+1} + a^*_i \delta_{j,i+1}.
\end{equation}
We require that $C$ describes a pure state, i.e.\ $C^2 = C$, which implies
\begin{equation}
    \begin{cases}
        1/4 + |a_j|^2 + |a_{j+1}|^2 = 1/2, \\
        a_j a_{j+1} = 0.
    \end{cases}
\end{equation}
We solve these two conditions by imposing $a_{2l} = 0$ and $a_{2l-1} = e^{i \varphi_l} /2$. We require also $\langle n_k\rangle = 1/2$ for any $k$
\begin{equation}
\label{eq:nkcond}
\begin{split}
    \langle n_k\rangle &= \frac{1}{L} \sum_{j,l} e^{-i k (j-l)} C_{jl} = \frac{1}{2} + \frac{1}{L} \sum_{j} (e^{-ik} a_j + e^{ik} a^*_j) \\
    &= \frac{1}{2} + \frac{1}{ L} \sum_{l = 1}^{L/2} \cos(k+\varphi_l).
    \end{split}
\end{equation}
Assuming $L$ to be even, any choice of $\varphi_l = \phi + \pi  \left(1-(-1)^l \right)/2$ makes the sum in the last term of Eq.~\eqref{eq:nkcond} vanish. Summarizing the result so far, we have
\begin{equation}
\label{eq:app_Ctridiag}
    C_{jj}= \frac{1}{2}, \qquad C_{2l,2l-1} = C^*_{2l-1,2l}=\frac{1}{2}(-1)^l e^{i \phi}
\end{equation}

Finally, we need to require that $C$ is not $\mathcal{T}-$invariant. Using Eq.~\eqref{eq:Tau_def} we can show that, if $C$ describes a state $\mathcal{T}-$invariant then, $\mathfrak{Re} C_{j,j-1} = 0$, to see this note that, in terms of complex fermions
\begin{equation}
    \mathcal{T} c_j \mathcal{T}^{-1} = (-1)^{j+1} c_j.
\end{equation}
The transformation of the correlation under $\mathcal{T}$ is then
\begin{equation}
\label{eq:appTC}
    C_{ij} \xrightarrow[\mathcal{T}]{} (-1)^{i+j} C^*_{ij},
    \end{equation}
where the complex conjugation appears because of the antiunitarity of $\mathcal{T}$.
Equation~\eqref{eq:appTC} implies that, for a $\mathcal{T}$-invariant $C$, $\mathfrak{Re} C_{j,j+(2k+1)} = 0$ and that $\mathfrak{Im} C_{j,j+2k} = 0$ for integer $k$.
As a consequence, any choice $\phi$ such that $\phi \neq \pm \pi/2$ corresponds to a pure Gaussian state locally relaxing to the infinite temperature state not invariant under $\mathcal{T}$.

The same construction can straightforwardly be extended to initial correlation matrix with longer-range correlations.

\section{Born-typical bath charge variance}
\label{app:variance}

In this appendix, we analyze the Born-typical properties of the post-measurement bath charge distribution. We begin with the explicit example of the particle number $N_B$. As derived in Eq.~\eqref{eq:pnbz} of the main text, we have
\begin{equation}
\label{eq:app_pnbz}
  p_{\text{out}}(N_B|\bm{z}) 
  = \frac{1}{2^{b_{\bm{z}}}} \binom{b_{\bm{z}}}{\frac{N_B - (L_B/2-b_{\bm{z}})}{2}},
\end{equation}
where $b_{\bm{z}}$ is the number of measured bonds with even parity. The distribution \eqref{eq:app_pnbz} has $\langle N_B\rangle_{\bm{z}} = L_B/2$ and $\text{Var}_{\bm{z}}(N_B) = b_{\bm{z}}$. To characterize the Born-typical variance we thus need to study the properties of $b_{\bm{z}}$ under the Born distribution. By definition, we can write $b_{\bm{z}} = \sum_{j=1}^{L_B/2} \eta_j(\bm{z})$, where $\eta_j(\bm{z}) = (1+\langle P_j\rangle_{\bm{z}})/2$. At a fixed time $t$ along the quench
\begin{equation}
\label{eq:bornbz}
    \exB[b_{\bm{z}}] = \sum_{j = 1}^{L_B/2}\frac{1+\langle P_j\rangle_t}{2} = \frac{L_B}{4}+\frac{1}{2} \sum_{j} \langle P_j\rangle_t.
\end{equation}
Assuming local relaxation to the infinite temperature state, we obtain that 
\begin{equation}
\label{eq:lim_bornbz}
\lim_{t\to\infty}\lim_{L_B\to\infty} \frac{\exB[b_{\bm{z}}]}{L_B} = \frac{1}{4}.    
\end{equation}
We now consider the Born variance of $b_{\bm{z}}$
\begin{equation}
    \text{Var}_{\text{Born}}[b_{\bm{z}}] = \sum_{j,k} \left(\exB[\eta_j(\bm{z}) \eta_k(\bm{z})]- \exB[\eta_j(\bm{z})] \exB[\eta_k(\bm{z})]\right),
\end{equation}
using the Born rule
\begin{equation}
    \text{Var}_{\text{Born}}[b_{\bm{z}}] = \sum_{j,k} \left(\langle \eta_j \eta_k\rangle_t- \langle \eta_j\rangle_t\langle\eta_k\rangle_t\right),
\end{equation}
where $\eta_j$ without $\bm{z}$ dependence is intended as the operator $\eta_j = (\mathbb{I} + P_j)/2$. Because we are fixing a certain time $t$, connected correlations are developed only on a scale fixed by $t$ and the following bound is satisfied
\begin{equation}
    \text{Var}_{\text{Born}}[b_{\bm{z}}] \leq f(t) L_B.
\end{equation}
We use now Chebyshev's inequality~\cite{durrett-19}, obtaining 
\begin{equation}
\text{Prob}_{\text{Born}}\left(\left|\frac{b_{\bm{z}}}{L_B} - \frac{\exB[b_{\bm{z}}]}{L_B}\right|\geq \varepsilon\right) \leq \frac{f(t)}{\varepsilon^2 L_B}, 
\end{equation}
performing the limit in the correct order
\begin{equation}
    \lim_{t \to \infty} \lim_{L_B \to\infty} \text{Prob}_{\text{Born}} \left(\left|\frac{b_{\bm{z}}}{L_B}- \frac{1}{4}\right|\geq \varepsilon\right) = 0,
\end{equation}
which implies that $b_{\bm{z}}$ will concentrate on the average value $\exB[b_{\bm{z}}] = L_B/4$, indicating that $\text{Var}_{\bm{z}}(N_B) \simeq L_B/4$ for typical Born trajectories as stated in the main text.

The same logic can be used to derive the typical extensive variance of the post-measurement bath distribution of a generic quadratic conserved charge $Q^B_{\bm{u}}$, plugging the explicit form of $\Gamma_{B,\bm{z}}$ (Eq.~\eqref{eq:GBz}) in Eq.~\eqref{eq:muzVarz}
\begin{equation}
\label{eq:varz2}
    \text{Var}_{\bm{z}}(Q^B_{\bm{u}}) = -\frac{1}{8} \left[ \text{Tr}[X_{\bm{u}}^2] + \sum_{\alpha,\beta} z_\alpha z_\beta \text{Tr}[E_{\alpha} X_{\bm{u}} E_{\beta} X_{\bm{u}}]\right],
\end{equation}
notice that the trace $\text{Tr}[E_{\alpha} X_{\bm{u}} E_{\beta} X_{\bm{u}}] = 0$ whenever $\alpha$ and $\beta$ correspond to Majorana bilinears living on bonds at a distance larger than $L_A$ (maximum range of $X_{\bm{u}}$). For this reason, the double sum in Eq.~\eqref{eq:varz2} contains only $O(L_B)$ terms and we can decompose the variance as 
\begin{equation}
    \text{Var}_{\bm{z}}(Q^B_{\bm{u}}) \simeq \sum_j q_j(\bm{z}),
\end{equation}
where each $q_j(\bm{z})$ only depends on outcomes in a neighborhood containing $O(L_A)$ sites. We can now repeat exactly the same argument as for the particle number, obtaining 
\begin{equation}
    \lim_{t \to \infty} \lim_{L_B \to \infty} \frac{\exB\left[\text{Var}_{\bm{z}}(Q^B_{\bm{u}})\right]}{L_B} = c_{\bm{u}}.
\end{equation}
For any nonzero linear combination $\bm{u}$, the absence of exactly revealed conserved charges established in the main text implies $c_{\bm{u}} >0$.
Once the Born-average is proven to be extensive, the rest of the derivation is the same as the one for $b_{\bm{z}}$, thus establishing the extensive bath charge variance for Born-typical trajectories.

Notice that this result also applies to the finite-temperature case, the only point in which the infinite temperature assumption enters is in determining the Born-averaged density variance, see e.g. Eq.~\eqref{eq:bornbz} and Eq.~\eqref{eq:lim_bornbz}, while the scaling behaviour in $L_B$ is unaffected.

\section{$\sqrt{\rho_{\mathrm{GGE},A}}-$distortion of a Gaussian state}
\label{app:rhodist}
In this appendix, we derive the formula in Eq.~\eqref{eq:gammadist} describing the effect of the $\sqrt{\rho_{\mathrm{GGE},A}}-$distortion of a Gaussian state on the correlation matrix. 
Our goal is to determine the correlation matrix $\Gamma'$ of the state
\begin{equation}
    \ket{\psi_{\Gamma'}} = \frac{\sqrt{\rho_{\mathrm{GGE},A}} \ket{\phi_{\Gamma}}}{\sqrt{\bra{\phi_{\Gamma}} \rho_{\mathrm{GGE},A}\ket{\phi_{\Gamma}}}},
\end{equation}
knowing the correlation matrices $\Gamma_{\mathrm{GGE},A}$ and $\Gamma$ of $\rho_{\mathrm{GGE},A}$ and $\ket{\phi_{\Gamma}}$ respectively. In the following, to simplify the notation, we denote by $\trho$ the distortion $\rho_{\mathrm{GGE},A}$ and by $\widetilde{\Gamma}$ its correlation matrix.

To determine $\Gamma'$, we can interpret the operation $\sqrt{\trho} \ket{\phi_{\Gamma}}$ as the projection onto the pure state obtained from the purification of the mixed state $\trho$, for which the associated transformation on the correlation matrix transformation is known, see Eq.~\eqref{eq:proj_formula}. More precisely, we double the degrees of freedom in $A$ introducing the copy $\overline{A}$. The purification of $\trho$, denoted by $\ket{\sqrt{\trho}}\!\rangle$, is defined as 
\begin{equation}
    \ket{\sqrt{\trho}} \!\rangle = \sum_n \sqrt{\tilde{p}_n} \ket{n}_{A} \ket{n}_{\overline{A}},
\end{equation}
where $\{\ket{n}\}$ is the eigenstate basis of $\trho$. The purified state satisfies, by construction, the property $\text{Tr}_{\overline{A}}\left[\ket{\sqrt{\trho}}\!\rangle \langle\! \bra{\sqrt{\trho}}\right] = \trho$. Therefore, the correlation matrix $\Gamma_p$ of the purified state can be taken as
\begin{equation}
    \Gamma_p = \begin{pmatrix}
        \widetilde{\Gamma} & S\\
        -S^T & -\widetilde{\Gamma}
    \end{pmatrix}.
\end{equation}
Requiring the pure state condition $\Gamma_p^2 = - \mathbb{I}$ fixes the off-diagonal block $S$ to $S =\sqrt{\mathbb{I} + \widetilde{\Gamma}^2}$.
Let now $\ket{\phi}_{\overline{A}} = \sum_n \phi_n \ket{n}_{\overline{A}}$ be a pure state in $\overline{A}$, one has
\begin{equation}
\label{eq:app_proj}
    (\mathbb{I}_A \otimes
     \bra{\phi^*}) \ket{\sqrt{\trho}}\!\rangle = \sqrt{\trho} \ket{\phi},
\end{equation}
as can be explicitly derived, since 
\begin{equation}
     (\mathbb{I}_A \otimes
     \bra{\phi^*}) \ket{\sqrt{\trho}}\!\rangle = \sum \phi_n \sqrt{\Tilde{p}_n} \ket{n}_A,
\end{equation}
which is precisely the decomposition of $\sqrt{\trho} \ket{\phi}$ in the eigenbasis of $\trho$. Applying Eq.~\eqref{eq:app_proj}, we can directly use the Gaussian projection formula (see Eq.~\eqref{eq:proj_formula}) by projecting $\overline{A}$ on the Gaussian state $\ket{\phi_{\Gamma}^*}$ with covariance $-\Gamma$, obtaining
\begin{equation}
\label{eq:app_projformula}
    \Gamma' = \widetilde{\Gamma} + S (-\widetilde{\Gamma} - \Gamma)^{-1} S.
\end{equation}
Using $\Gamma^{-1} = - \Gamma$ since $\ket{\phi_{\Gamma}}$ is pure, we have $(-\widetilde{\Gamma} - \Gamma)^{-1} =  \Gamma (\mathbb{I}-\widetilde{\Gamma} \Gamma)^{-1}$, which plugged into Eq.~\eqref{eq:app_projformula} returns the formula~\eqref{eq:gammadist} of the main text.

\end{document}